\documentclass[12pt]{article}

\usepackage{color,epsfig}
\usepackage{graphicx}

\definecolor{purple}{rgb}{0.6,0,0.2}
\definecolor{violet}{rgb}{0.4,0,0.6}
\definecolor{vert}{rgb}{0,0.4,0.2}
\definecolor{navy}{rgb}{0.0,0.0,0.4}
\def\colbrun#1{\textcolor[named]{Brown}{#1}}

\def\spose#1{\hbox to 0pt{#1\hss}}\def\lta{\mathrel{\spose{\lower 3pt\hbox
{$\mathchar"218$}}\raise 2.0pt\hbox{$\mathchar"13C$}}}  \def\gta{\mathrel
{\spose{\lower 3pt\hbox{$\mathchar"218$}}\raise 2.0pt\hbox{$\mathchar"13E$}}}

\def\be{\begin{equation}}
\def\fe{\end{equation}}

\def\mm{{\color{red}m}}
\def\pprime{{\color{red}\prime}}
\def\llam{{\color{red}\lambda}}
\def\caL{{\color{red}{\cal L}}}
\def\calP{{\color{red}{\cal P}}}
\def\Thheta{{\color{red}\Theta}}

\def\Ttheta{{\color{violet}{\mit \Theta}}}
\def\Ssigma{{\color{violet}{\mit \Sigma}}}
\def\XX{{\color{violet}X}}
\def\calX{{\color{violet}{\cal X}}}
\def\calY{{\color{violet}{\cal Y}}}
\def\calZ{{\color{violet}{\cal Z}}}

\def\TT{{\color{red}T}}
\def\UU{{\color{red}U}}
\def\WW{{\color{red}W}}
\def\VV{{\color{red}V}}
\def\HH{{\color{red}H}}
\def\SS{{\color{red}S}}
\def\PP{{\color{red}P}}
\def\KK{{\color{red}{\cal K}}}
\def\rrho{{\color{red}\rho}}
\def\AA{{\color{red}A}}
\def\QQ{{\color{red}Q}}
\def\GG{{\color{red}G}}
\def\KG{{\color{red}K}}
\def\Kdel{{\color{red}\delta}}
\def\eeta{{\color{red}\eta}}
\def\ZZ{{\color{red}Z}}

\def\aalpha{{\color{purple}\alpha}}
\def\ddelta{{\color{purple}\delta}}
\def\arepsilon{{\color{purple}\varepsilon}}

\def\bbeta{{\color{blue}\beta}}
\def\ww{{\color{blue} w}}

\def\delt{{\colbrun\delta}}

\def\mmu{{\color{blue}\mu}}
\def\Upsi{{\color{violet}\Upsilon}}
\def\Pssi{{\color{violet}\Psi}}
\def\Phhi{{\color{violet}\Phi}}

\def\xsi{{\color{violet}\xi}}

\def\phhi{{\color{violet}\phi}}
\def\thheta{{\color{violet}\theta}} 
\def\chhi{{\color{violet}\chi}}
\def\pssi{{\color{violet}\psi}}
\def\ssigma{{\color{violet}\sigma}}
\def\zzeta{{\color{violet}\zeta}}

\def\ge{{\color{blue}g}}
\def\vv{{\color{blue}v}}
\def\ss{{\color{blue}s}}

\def\elle{{\color{vert}\ell}}
\def\nn{{\color{vert}n}}
\def\nnu{{\color{vert}\nu}}

\def\Rh{{\color{blue}R}_{\rm H}}
\def\aaleph{{\color{vert}\aleph}}
\def\tt{{\color{blue}t}}
\def\aa{{\color{blue}a}}
\def\xxi{{\color{vert}R}}

\begin{document}

\begin{center}

\colbrun{\em {\it Contribution for Peyresq Physics Meeting:} 
{\it  Macro and Micro Structure of Spacetime}.}
\\[0.6 cm]

 \textcolor{red}{\Large Stability of winding cosmic 
wall lattices \\ with X type junctions
}    \\[1.0cm]
 {\bf Brandon Carter} \\ [0.5cm]
{ LuTh, Observatoire Paris,  92195 Meudon, France. }
\\[0.6cm]
 \color{vert} December, 2007.  \\

\end{center}

\vskip 1 cm

\noindent
{\bf Abstract.} This work confirms the stability of a class of domain 
wall lattice models that can produce accelerated cosmological expansion, 
with pressure to density ratio $\ww=-1/3$ at early times, and with 
$\ww=-2/3$ at late times when the lattice scale becomes large compared to 
the wall thickness. For walls of tension $\TT_{\rm I}$, the relevant X type 
junctions could be unstable (for a sufficiently acute intersection angle 
$\aalpha$) against separation into a pair of Y type junctions joined by a 
compound wall, only if the tension $\TT_{\rm I\!I}$ of the latter were 
less than $2\TT_{\rm I}$ (and for an approximately right-angled 
intersection if it were less that $\sqrt{2}\,\TT_{\rm I}$) which can not 
occur in the class considered here. In an extensive category of 
multicomponent scalar field models of forced harmonic (linear or 
non-linear) type it is shown how the relevant tension -- which is the 
same as the surface energy density $\UU$ of the wall -- can be calculated 
as the minimum (geodesic) distance between the relevant vacuum states as 
measured on the space of field values $\Phhi^i$ using a positive definite 
(Riemannian) energy metric ${\rm d}\UU^2=\tilde \GG_{ij}\,
{\rm d}\Phhi^i\,{\rm d}\Phhi^j$ that is obtained from the usual kinetic 
metric (which is flat for a model with ordinary linear kinetic part) by 
application of a conformal factor proportional to the relevant potential 
function $\VV$. For suitably periodic potential functions there will be 
corresponding periodic configurations -- with parallel walls characterised 
by incrementation of a winding number -- in which the condition for 
stability of large scale bunching modes is shown to be satisfied 
automatically. It is suggested that such a configuration -- with a 
lattice lengthscale comparable to intergalactic separation distances 
--  might have been produced by a late stage of cosmological inflation.

\vskip 1.cm
\vfill\eject

\noindent
{\bf 1. Introduction}
\medskip

This article is concerned with the question raised by  Bucher and 
Spergel \cite{BS98,BBS99} of whether some kind of cosmic domain wall lattice 
might account for the observed acceleration of the expansion of the universe,
not to mention subtle deviations from isotropy\cite{BM06,BM07}.
 More specifically, the question is whether the tendency of such lattices 
to evolve -- typically according to a scaling law \cite{BM06b} -- towards
a uniform vacuum solution can be ``frustrated'' in particular scenarios in which
the lattice is ultimately preserved  in a configuration that is effectively 
frozen with respect to comoving coordinates. A first rerequisite that must be 
satisfied by proposed candidate models \cite{C04,BCCM05,BCM05} for such a frozen 
state is of course that of stability, but even in cases for which this is 
satisfied there remains, as a further necessary condition for viability, the 
more difficult problem of attainability of the configuration in question 
from plausible initial conditions.

The first aim of the present work is to show how the existence of
absolutely stable frozen lattice configurations will be an automatic consequence 
of the existence of conserved topological winding numbers for an extensive class
of field models with multiply connected field configuration spaces. 
Although such models would seem to be capable of providing good agreement
with what we see now, this depends on the provision of initial conditions
for which -- contrary to what might be expected from naive symmetry 
considerations -- the relevant winding numbers are endowed with
non zero values. That nature might conceivably provide such symmetry violation --
perhaps for anthropic reasons -- is shown by the notorious example of
cosmic baryon number asymmetry. The need to invoke such a priori
symmetry violation does however diminish the attractivity of such
scenarios.

In the (not yet fully  satisfied) hope of getting round this drawback, the 
following work will be mainly concerned with the extension of the class of 
field models under consideration to a more general class in which, when 
sufficient energy is available, the fields will have access to a 
configuration space that is simply connected, but in which the energetically 
attainable field values will otherwise be effectively confined to a  
neighbourhood that is postulated to have  the non-trivial multiply connected 
kind of topology considered in the preceding paragraph. In scenarios based 
on such models, cooling from a thermally excited state can be expected to 
leave field configurations with winding numbers that will be effectively 
conserved by substantial energy barriers, and that may be left with non-vanishing 
values -- thus guaranteeing ``frustration'' -- over causally connected volumes 
that in an inflationary scenario could be larger that the visible universe.
However an (in the present context) undesired biproduct of this process
would be the formation of string type defects on which the domain walls 
resulting from the winding would terminate. It is easy to conceive ways in
which such undesired strings might be inflated away (as in the usual solution
of the monopole problem) but would seem that to do this without also
inflating away the desired wall lattice would again require recourse to fine 
tuning (and perhaps invocation of the anthropic principle) such as we were 
trying to avoid. The ultimate message of this article is that this
approach does not seem very promising at the present stage, but that it should
not yet be definitively excluded from consideration.

It is to be remarked that a more decisively negative opinion, namely that 
viable scenarios for a ``frustrated'' lattice cannot exist, has been vehemently 
advocated in  recent  articles by Avelino {\it et al.} \cite{AMMM06,AMMM06b}.  
(Since they went so far as to claim that the viability of my proposed pentavac 
example\cite{C04} was ``easily''  ruled out by their numerical work, despite the 
fact that the latter was confined to a limited  -- apparently inappropriate --  
part of the relevant parameter space, the present article will therefore give 
particular attention to the provision of analytic reasonning showing that 
this pentavac model does indeed provide  lattices with the required stability 
properties provided the relevent parameters are chosen within the appropriate 
range as evaluated in an appendix.)  In more recent writing \cite{AMMM07} these 
authors have conceded that our examples \cite{C04,BCCM05,BCM05} do show how 
``one can build (purely by hand) special lattices that would be locally stable 
against small perturbations''.  They nevertheless maintain their 
{\it no frustration conjecture} to the effect that ``no such configurations 
are expected to ever emerge from any realistic cosmological phase transition''. 
I would suggest that a more justifiable conclusion (from their own and other 
concordant work on the tendency towards scaling behaviour \cite{BM06b}) would 
be obtainable by substituting the qualification ``easily numerically simulable'' 
in place of their adjective ``realistic'', but I agree that the ``frustrated''
examples in this and the preceeding work \cite{C04,BCCM05,BCM05} can quite
fairly be criticised as artificially contrived. What should not be overlooked 
however is that one can also criticise wild flowers as having been artificially 
contrived to attract bees: the point is that artificially contrived results are 
sometimes obtainable by Darwinian or anthropic selection mechanisms in a manner 
that is undeniably natural and ``realistic'', albeit far beyond the scope of easy 
numerical simulation.

\begin{figure}
\centering
\epsfig{figure=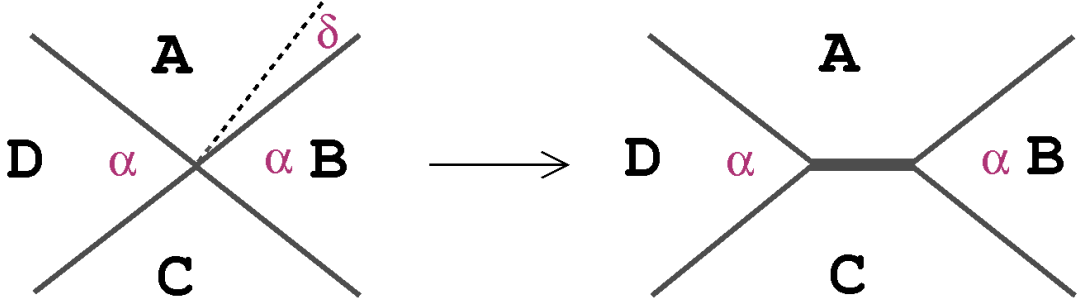, width=15 cm}
\caption{Transition from X junction (deviating from orthogonality
by angle $\ddelta$) to neighbouring pair of Y junctions 
(with complementary intersection angle $\aalpha$) for walls 
between vacuum domains labelled A, B, C, D, using heavy line to 
indicate compound wall.}
\label{XtoY}
\end{figure}

Before proceding to what is new, this article will start by recapitulating some 
noteworthy conclusions from the preceding work \cite{C04,BCCM05,BCM05} in which 
attention was drawn to the important qualtative distinction between the X type 
of (crossover) junction that is more favorable  for stability of the lattice, 
and  the Y type of junction, in which there is no freedom of adjustment in the 
equilibrium angle of intersection (which must be $\pi/3$ if the wall tensions 
are all equal). For an X type crossover the most symmetric possibility is a right 
angle intersection, but for opposing pairs with equal tension equilibrium will 
still possible when there is a positive  deviation $\delta$ so that the walls 
meet at an acute angle
{\be \aalpha=\pi/2-\ddelta \, ,\label{delt}\fe} 
It has however been emphasized \cite{AMMM06} that in order to contribute to a 
stable lattice such an X type equilibrium must be stable against decomposition 
into a pair of Y type junctions (see Fig. \ref{XtoY}), a requirement that was 
not explicitly checked in the particular toy field models I suggested \cite{C04} 
as examples in my original discussion of this subject, and that has been called 
into doubt \cite{AMMM06} in the particular case of what will be referred to here 
as the pentavac doublet model.

In order to address the lattice stability question, a preliminary task of the 
present work will to provide a simple general criterion for the stability of 
such an X type equilibrium junction. A strategy for testing this criterion will 
then be provided for an extensive class of forced harmonic field models, 
including the pentavac doublet model \cite{C04}, for which it is confirmed that 
(contrary to the doubts that have been expressed \cite{AMMM06,AMMM06b}) the 
stability condition is indeed satisfied. The main part of this work is concerned 
with the related but more delicate issue of stability against bunching of parallel 
walls, which will be dealt with in the same framework.

Although -- like many other possibilities such as the monovac triplet model 
developed at the end of this article -- the pentavac doublet model can provide a 
regular lattice that is stable, it should be mentioned that it is nevertheless 
unsatisfactory from the point of view of the purpose that motivated its 
introduction, which was to provide a random lattice of the kind to which my 
(still unproven) five colour conjecture was concerned\cite{C04}. The special 
feature of the pentavac doublet model (see Fig. \ref{ABCD} at the end) is the 
admission of simple domain walls between any of the 10 pairs that can be chosen 
from its 5 distinct but equivalent vacua, and the admission of X type 
crossovers involving any of the 5 possible combinations of 4 distinct vacuua. 
However, for each such combination, the pentavac model allows only one of the 
3 mathematically conceivable ways of choosing the dagonally opposing pairs, 
whereas all of them would be needed for a random solution of the five 
colour problem.

This limitation on all the cases investigated so far -- namely that 
they provide lattices that can be stable only when sufficiently regular, 
not random, and hence that they seem to depend on prerequisite ``tuning'' 
of the universe -- has considerably reduced the attractivity of such domain 
wall models in comparison, for example, with the simple hypothesis of an 
appropriately ``tuned'' cosmological constant. 

Despite this limitation, wall scenarios of the periodic lattice type 
considered here remain viable in principle,  as a possibility that 
should perhaps be taken more seriously, particularly \cite{BM06,BM07} if 
the evidence for cosmological anisotropy \cite{RWULL04} is confirmed. 
The systems proposed for investigation here are of a kind that would 
arise from multiplet generalisations of the much studied Peccei-Quinn 
singlet model, which produces axionic walls whose potentially important
cosmological consequences have been considered by Khlopov and coauthors
\cite{Khlopov85,Khlopov98,Khlopov04}. In systems of this kind, the 
collective stability of the walls depends on their feature of having 
topological winding numbers, which need to add up coherently with the 
same sign. The problem with this is that in a random system (such as 
would be expected from a Kibble type symmetry breaking mechanism 
\cite{GH03,APV04,OMA05}) one would expect to obtain roughly equal 
numbers of positive and negative windings (which in the long run would 
undergo mutual destruction, leaving a residue of string anomalies).  

In comparison with the (unavoidable) problem of accounting for the 
observed (but still mysterious \cite{BKY05}) preponderance of ordinary 
matter over antimatter, the analogous, but numerically less extreme, 
problem of accounting for the required preponderance (on a sufficiently 
large scale) of positive over negative walls appears to be less 
intractable. The reasonning developped below suggests that it may be 
soluble on the basis of statistical fluctuations in the framework of 
suitable inflationary models -- albeit with the help of special 
parameter tuning such was already inherent in such models. 
The basic idea is that -- assuming the strings that might have been 
formed in a  very early high (e.g. GUT) energy transition were 
subsequently inflated away -- reheating could have produced walls
associated with more moderate (e.g. electroweak) energy scales, that
(due to statistical fluctuations) need not have exactly cancelled 
themselves out in a localised volume, but could have left a residue 
with coherent winding on a sufficiently large (cosmologically 
significant) scale.

\bigskip\noindent
{\bf 2. Simple and compound wall configurations in models with
X type junctions.}
\medskip

The recently raised issue\cite{AMMM06} of possible instability of X type 
junctions arises in bosonic field models of the kind I suggested as 
candidates for providing domain wall domain wall lattices, in which the 
essential feature -- as illustrated in Fig. \ref{ABCD} -- is the 
existence in the space of classical field values of neighbouring subsets 
of four equivalent discrete vacuua -- meaning minima of the potential 
energy $\VV$ -- separated by four ridges of higher energy that meet in a 
cross shaped configuration at a peak of even higher energy. Using the 
letters A, B, C, D, to label the energy minima in cyclic order, as one 
goes round the peak of $\VV$ in configuration space, entails the 
concomitant notation AB, BC, CD, DA to label the ridges that separate 
them.

For each such ridge in the space of field values, there will be a 
corresponding flat domain wall equilibrium state in ordinary spacetime: 
for example AB will designate the -- energy minimising -- domain wall 
state specifiable as a function of a single cartesian cooordinate, $x$, 
say by the condition that it tends to the field configuration A as 
$x\rightarrow -\infty$ and to field configuration B as 
$x\rightarrow +\infty$, with maximum energy density at $x=0$. Since 
there is nothing to break the Lorentz invariance in directions parallel 
to the $x=0$ plane, such a wall will be a Dirac type brane, with 
isotropic tension, $\TT_{\rm I}$ equal to its surface energy density. 

As well as such simple branes associated with the field space energy 
ridges  AB, BC, CD, DA, there may also be compound wall configurations 
separating non adjacent vacuum pairs, for which the possible combinations 
are AC and BD. In order to be stable against splitting into the relevant 
pair of simple walls (for example in order for the compond wall AC to 
be stable against splitting into the separate simple walls AB and BC) 
the field model should evidently be such that the energy density (and 
tension) $\TT_{\rm I\!I}$ say of the compound wall configuration -- 
if it exists -- is less  than  twice that of a single wall:
{\be \TT_{\rm I\! I} < 2\TT_{\rm I} \label{1}\, .\label{comstab}\fe}

\bigskip\noindent
{\bf 3. Criterion for  stability of an X type junction}
\medskip

In a model of the kind considered in the previous section, an 
ordinary X type junction between four vacuum domains A,B,C,D, 
consists of a string like locus of intersection through which a 
simple wall of type DC continues as a simple wall of type AB, while 
a simple wall of type AD continues as a simple wall of type BC.
Such an X type junction will always be stable if the model does not
admit any compound wall configurations satisfying (\ref{1}).

The possibility \cite{AMMM06} of instability with respect to 
decomposition into a pair of Y type junctions will however arise if 
the model is such as to admit compound walls satisfying the condition 
(\ref{1}).  If the simple walls AB and BC meet at an acute angle
$\aalpha\ ,$ the decomposition in question would create a Y junction
joining them to a compound wall of type AC, while -- as shown in 
Fig. \ref{XtoY} -- at the other end of the double wall there would be 
another Y type junction on which the simple walls AD and DC would meet 
at the same acute angle $\aalpha$. 
The original X type junction will be unstable if and only if the 
double wall of type AC is too weak to prevent the separation of the 
two Y junctions from increasing, that is to say if and only if
{\be \TT_{\rm I\!I} <2\TT_{\rm I}\, {\rm cos}\,\{\aalpha/2\}\, ,\fe}
a condition that is evidently stronger than (\ref{1})
This is expressible the other way round as the condition 
that the X junction will be stable if and only if the acute angle 
$\alpha$  is such that
{\be {\rm cos}\,\{\aalpha/2\}<\frac{\TT_{\rm I\!I}}{2\TT_{\rm I}} 
\, .\label{crita}\fe}

\begin{figure}
\centering
\epsfig{figure=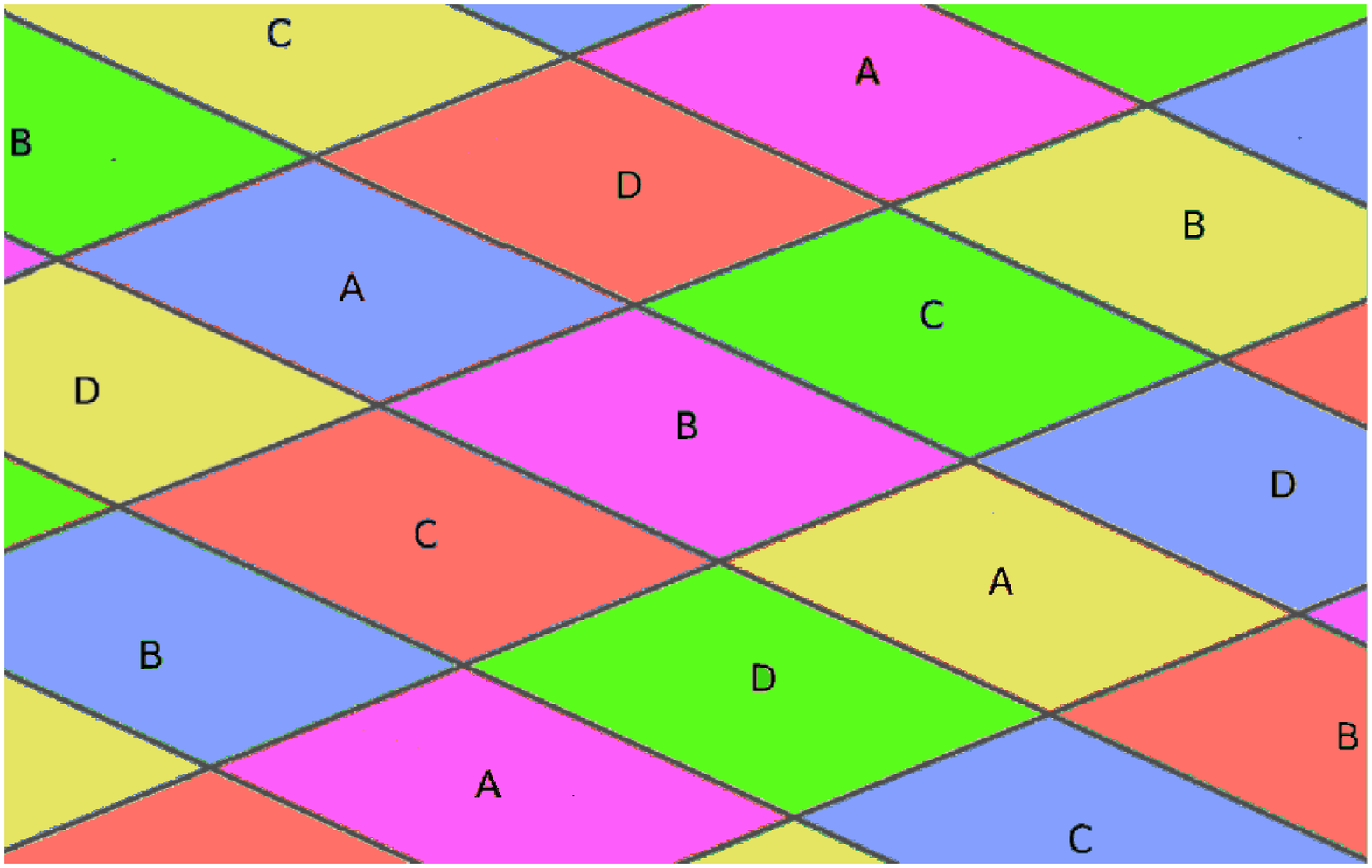, width=12 cm}
\caption{Locally stable periodic  lattice satifying stability 
condition (\ref{uncstab}) or (\ref{ancon}) with subcritical deformation 
angle $\ddelta$ for X junctions between walls separating  vacuum 
domains that might be identical, as in monovac winding models, but 
that might also for example be of 4 types as indicated by the
labels A, B, C, D, or even of 5 types as indicated by the coloring.
\bigskip}
\label{Xjunlat}
\epsfig{figure=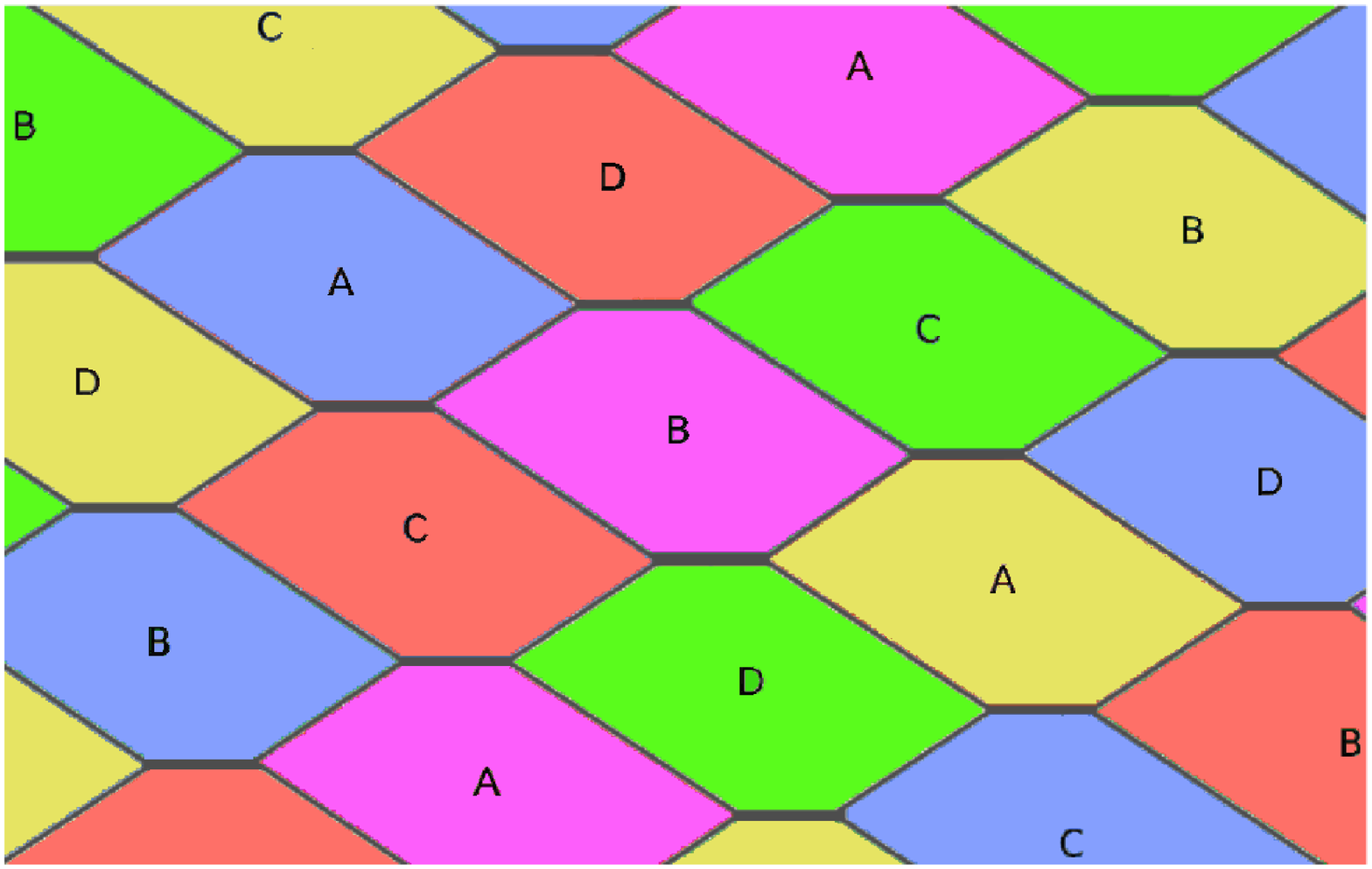, width=12 cm}
\caption{Periodic lattice for same kinds of model with same global 
(phase winding) boundary conditions and deformation angle 
$\ddelta$ as in Figure \ref{Xjunlat} but in regime 
(\ref{simwall}) with smaller critical value $\ddelta_{\rm c}<\ddelta$ 
so that junctions will be of Y type, with critical intersection angle 
$\aalpha_{\rm c}=\pi/2-\ddelta_{\rm c}\, .$ This lowers energy 
density by factor ${\rm cos}\{(\ddelta-\ddelta_{\rm c})/2\}\, ,$ 
giving configuration whose stability against local perturbations 
is marginal.}
\label{Yjunlat}
\end{figure}

\begin{figure}
\centering
\epsfig{figure=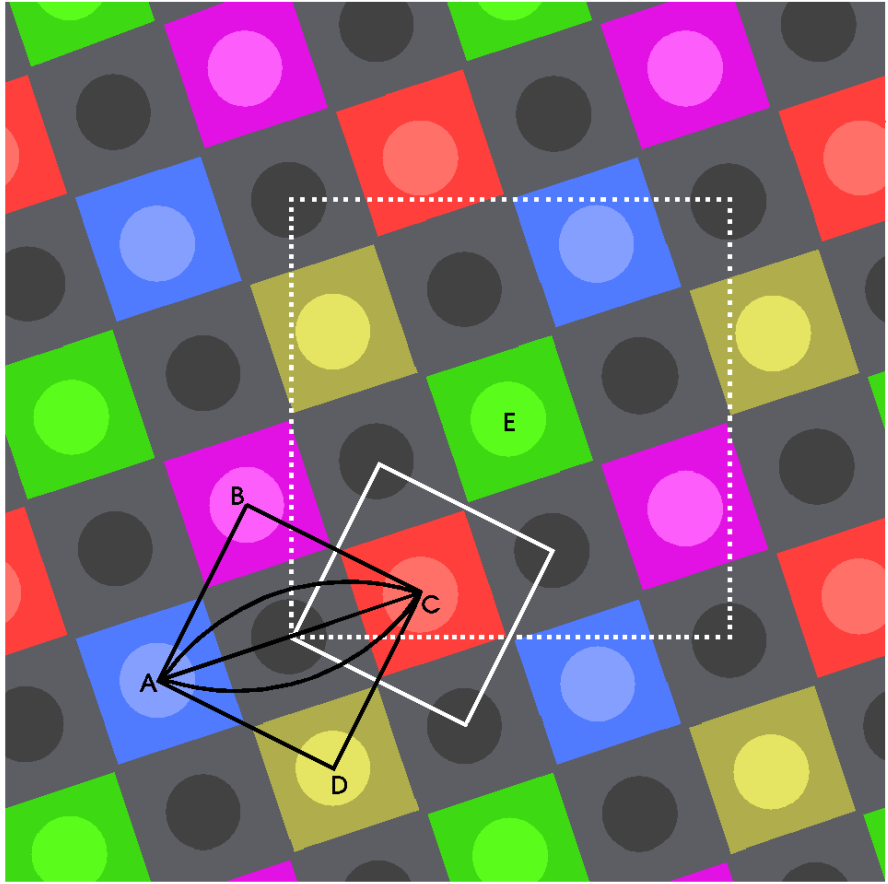, width=12 cm}
\caption{Contours of the potential $\VV$, using darkest shading at maxima 
and bright colours to distinguish distinct minima -- representing 
vacuua --  labelled A, B, C, D, E, showing the approximate location of 
some alternative paths from A to C, in plot of $\phhi$ against $\pssi$ 
representing  a (fourfold) periodic covering of toroidal field 
configuration space of pentavac doublet model with flat kinetic metric 
$\GG_{ij}\propto\Kdel_{ij}$ and energy metric $\tilde \GG_{ij}=2\VV 
\GG_{ij}$. The large white dotted square contains the single covering 
range $0\leq\phhi\leq 2\pi ,$ $0\leq \pssi\leq2\pi ,$
and the small white square contains the range $0\leq \thheta
\leq 2\pi ,$ $0\leq \chhi\leq 2\pi$ that would provide a
single covering for the corresponding monovac doublet model,
for which (as would be the shown by a black and white
printout of this figure) the five kinds of vacuum colour 
A, B, C, D, E would cease to be distinguishable. }
\label{ABCD}
\end{figure}

Since the range of possible values of the intersection angle is
given by $0<\aalpha\leq\pi/2\,  ,$ which is equivalent to
$1> {\rm cos}\,\{\aalpha/2\}\geq 1/\sqrt 2 \, ,$
it follows firstly that, in the regime for which 
{\be \TT_{\rm I\!I}\geq {2}\, \TT_{\rm I}\, ,\label{uncstab}\fe}
(so that the the criterion (\ref{comstab}) for local stability 
of an individual compound wall will fail)
the stability condition (\ref{crita}) will always be satisfied.
 Secondly, in the regime for which 
{\be2\TT_{\rm I}> \TT_{\rm I\!I}>\sqrt{2}\, \TT_{\rm I}\, ,
\label{simwall}\fe} there will be a critical crossing angle,
given by
{\be {\rm cos}\,\aalpha_{\rm c}={\rm sin}\,\ddelta_{\rm c}=
\frac{\TT_{\rm I\!I}^{\ 2}}{2 \TT_{\rm I}^{\,2}}-1\, ,
\label{ancrit}\fe}
such that the X junction stability condition  (\ref{crita}) will
hold if and only if the actual crossing angle satisfies
the criterion expressible in the equivalent forms
{\be \aalpha>\aalpha_{\rm c}\, ,\hskip 1 cm
\ddelta<\ddelta_{\rm c}\, ,\label{ancon}\fe}
in which case, as when (\ref{uncstab}) holds, it will be
possible to construct a stable periodic X-junction lattice 
of the kind illustrated in Figure \ref{Xjunlat}. Alternatively,
if the crossing angle were too acute for (\ref{ancon}) to
hold,  the same global boundary conditions could be satisfied
by a periodic Y junction lattice of the kind illustrated
in  Figure \ref{Yjunlat}, but  -- as in previously discussed
examples \cite{C04,BCCM05,BCM05}  -- the stability of such a 
configuration with respect
to local perturbations would be marginal.
Finally, if
{\be \TT_{\rm I\!I}\leq\sqrt{2}\, \TT_{\rm I}\, .\fe}
it is evident that the condition (\ref{crita}) for
 X junction stability will never 
be able to hold at all for any value of the crossing angle.

\bigskip\noindent
{\bf 4. Scalar field models of forced harmonic type}
\medskip

The toy bosonic field models we have been considering belong to a 
rather extensive category that is describable as being of forced 
harmonic type. This means that the independent scalar field 
components, $\Phhi^i$ say, are to be considered as coordinates on 
a manifold characterised by a flat or curved Riemannian 
(positive definite) metric
{\be {\rm d}\llam^2=\GG_{ij}\,{\rm d}\Phhi^i\,{\rm d}\Phhi^j
\, ,\label{lam}\fe}
and by a scalar forcing potential $\VV$, in terms of which the relevant 
Lagrangian density in ordinary spacetime, with pseudo Riemannian metric
${\rm d} \ss^2=\ge_{\mu\nu}\,{\rm d}x^\mu\,{\rm d}x^\nu$
takes the standard form
{\be {\caL}=-\frac{_1}{^2}\GG_{ij}(\nabla^\nu\Phhi^i)\nabla_\nu\Phhi^j
-\VV\, .\label{Lag}\fe}
Generalised non-linear sigma models (of the kind whose properties were 
used by Bunting \cite{C86} for his proof of black hole uniqueness) are 
included as the subcategory for which the potential function $V$ is set 
to zero. Scalar field models of the ordinary linear and non-linear kinds 
that are more commonly considered \cite{ABGT03} in physics are included 
as the subcategory for which the forcing potential is a variable function, 
$\VV\{\Phhi\}$, but for which the kinetic metric (\ref{lam}) on the space 
of field values is flat, so in such a kinetically linear model
 it will be possible to scale the field variables in such a way that the 
metric components will be given simply by the Kronecker unit matrix 
as $\GG_{ij}=\delta_{ij}$.

In a flat static domain wall state of the kind considered in the 
preceding section, for which the fields depend only on the single 
cartesian space coordinate $x$, the surface energy tension $ \TT$ will be 
the same as the integral along the $x$ axis of the energy density, as 
given in the limit $\elle\rightarrow\infty$ by the prescription
{\be \TT=\UU\{\infty\}\, , \label{en}\fe}
where
{\be \UU\{\elle\}=\int_{-\elle/2}^{\elle/2} \left(\frac{_1}{^2}\GG_{ij}
\frac{{\rm d}\Phhi^i}{{\rm d}x}\frac{{\rm d}\Phhi^j}{{\rm d}x}
+\VV\right){\rm d} x\, ,\label{varu}\fe}
in which (to get a finite result) it is to be understood that the 
potential $\VV$ has been adjusted if necessary, by the addition of a 
constant which will have no effect on the field equations obtained 
from (\ref{Lag}),  in such a way as to arrange for its minimum 
(vacuum) value to be zero, $\VV=0$.

Stable wall equilibrium states are characterised by the condition that 
the parametrised field manifold  trajectory (as specified by the 
functions $\Phhi^i\{x\}$) with endpoint at the relevant pair of vacuum 
position (where $\VV\{\Phhi\}$ is minimised) should be such that the 
energy integral (\ref{en}) should be a minimum.

Using a prime to denote derivation with respect to the parameter
$\llam$ measuring the distance in the field manifold, so that in 
particular, according to (\ref{lam}),  we shall have
{\be \GG_{ij}\Phhi^{i\pprime}\Phhi^{j\pprime}=1\, ,\fe}
it can be seen that  the integrand in (\ref{varu}) can be rewritten 
as
{\be {\rm d}\UU= \UU^\pprime\, {\rm d}\llam\, ,\label{emeas}\fe}
with
{\be \UU^\pprime=\frac{1}{2x^\pprime}+\VV x^\pprime\, ,\label{Uprime}\fe}
in terms of the rate of variation
{\be x^\pprime=\frac{{\rm d} x}{{\rm d} \llam} \, ,\fe}
of the space coordinate $x$ with respect to the field manifold
 distance given by (\ref{lam}). For given field values at the
extremities of some finite range $-\elle/2<x<\elle/2$, a corresponding 
equilibrium configuration will be characterised by minimisation
of the energy integral (\ref{en}) with the integrand given by
{\be \UU\{\elle\}=\int \UU^\pprime \,{\rm d}\llam\, ,\fe}
subject to the associated constraint
{\be \int x^\pprime\,{\rm d}\llam=\elle\, .\label{cons}\fe}
In terms of an appropriately adjusted value of a Lagrange multiplier
$\PP_{\!\perp}$ say, this condition will be equivalent to unrestricted 
minimisation of the corresponding enthalpy combination
{\be \HH\{\elle\}=\UU\{\elle\}+\PP_{\!\perp}\,\elle \, \fe}
which will be given by
{\be \HH=\int\left(\frac{1}{2x^\pprime}+(\VV+\PP_{\!\perp}) 
x^\pprime\right)\,{\rm d}\llam\, .\fe}
By considering the way this depends on $x^\pprime$ for a fixed
dependence of the fields on the parameter $\llam ,$ it can 
immediately be seen to be necessary for minimisation that the 
variation rate $x^\prime$ should satisfy the relation
{\be  x^{\pprime\, 2}=\frac{1}{2(\VV+\PP_{\!\perp})} \, ,\label{rest}\fe}
in which $\PP_{\!\perp}$ is interpretable as a constant of integration whose 
value can be seen to be that of the pressure in the direction orthogonal 
to the plane of the wall.  

The condition (\ref{rest}) can be used to eliminate the involvement of 
the ordinary space coordinate $x$, and to express the quantity that (for 
fixed field values at the endpoints of integration) has to be minimised 
in the form 
{\be \HH=\int \sqrt{2(\VV+\PP_{\!\perp})}\,\, {\rm d}\llam
\, ,\label{equmv}\fe}
in which only field manifold  variables are involved, and in which the 
fixed parameters $\elle$ and $\PP_{\!\perp}$ are not independent but 
according to (\ref{rest}) must be related by the consistency condition
{\be \elle=\int\frac{ {\rm d}\llam}{ \sqrt{2(\VV+\PP_{\!\perp})}}
\label{length}\, .\fe}
It can be seen from this that when the orthogonal pressure is varied
the corresponding variations of $\UU$ and $\elle$ will be related by
the formula
{\be \frac {{\rm d} \UU}{{\rm d}\elle}=-\PP_{!\perp}\, \label{pdef}\fe}
whereby the interpretation of $\PP_{\!\perp}$ as the relevant
orthogonal pressure is made obvious. 

In cases such as those of the compact, topologically non-trivial,
field manifolds to be considered in the next two sections, a 
solution for a finite value of $\elle$ may be extensible over the 
complete range of $x$ as a``lasagne'' type periodic configuration
of the kind whose (linear or non-linear) superposition will contitute 
the kind of cosmological lattice whose investigation is the ultimate 
motivation for this work. More specifically, the slab thickness $\ell$
will then be identifable  with the wavelength of the periodicity, so
that the corresponding longitudinal wavenumber density will be 
{\be \nnu_\perp=\frac{1}{\elle}= \frac{{\rm d}\PP_{\!\perp}}{{\rm d} 
\HH}\, ,\fe}
provided the endponts of the integration occur at successive maxima 
or successive minima of the potential as function of $x$. Thus in 
particular, for $\PP_{\!\perp}>0$, the distance $\elle$ will be the same as 
the wavelenth if the endpoints occur at successive vacuum states. 

In such an effectively one dimensional lasagne type configuration,
the condition for stability of the large scale averaged system against 
development of a bunching mode is the positivity the square 
$\vv_\perp^{\,2}$ of the orthogonal propagation speed, of 
perturbations of the mean density $\UU\nnu_\perp$ -- on length scales 
large compared with the wavelength $\elle$ -- as given by
{\be \vv_\perp^{\,2}=\frac{{\rm d}\PP_{\!\perp}}{{\rm d}(\UU\nnu_\perp)}
= -\frac{\elle}{\HH}\frac{{\rm d}\HH}{ {\rm d}\elle}\, .\label{vel}\fe}
That the requirement $\vv_\perp^{\,2}>0$ will indeed be satisfied
is shown by the expression
{\be \vv_\perp^{-2}=-\frac{\HH}{\elle^2}\frac{{\rm d}\elle}{{\rm d}
\PP_{\!\perp}}\, ,\hskip 1 cm -\frac{{\rm d}\elle}{{\rm d}\PP_{\!\perp}}
=\int \Big(2(\VV+\PP_{\!\perp})\Big)^{-3/2}{\rm d}\llam\, .\label{speed}\fe}

It can be seen (using the Schwarz inequality) that the velocity given by 
(\ref{speed}) can never exceed unity, meaning the speed of light. It will 
however approach this upper bound when $\PP_{\!\perp}$ becomes very large, 
so that the kinetic energy becomes much greater that the potential energy
(whose particular form will then be irrelevant) which is what happens
in the short wavelength limit $\elle\rightarrow 0  ,$ for which one 
will have
{\be \UU\sim \PP_{\!\perp}\elle\, ,\hskip 1 cm \PP_{\!\perp}\sim 
\frac{_1}{^2} (\llam/\elle)^2\, ,\hskip 1 cm \vv_\perp^{\,2}\sim 1
\, ,\label{hype}\fe} where $\llam$ is the integrated distance along the 
field space trajectory from one vacuum state to the next, as measured 
with respect to the kinetic metric (\ref{lam})

Our main concern here is not with the short wavelength limit,
but on the contrary with isolated wall configurations, which are 
obtained when the range of integration is unlimited, $\elle\rightarrow 
\infty$. This large separation condition requires that the integral 
(\ref{length}) should be divergent at the extremities of the trajectory 
in field space, namely the vacuum states characterising the domains in 
question, where the potential $\VV$ reaches its minimum. Since, for 
applicability of the tension formula (\ref{en}), it is to be understood 
that (by subtraction, if necessary, of any contribution from a 
cosmological constant that may be present) this minimum value of $\VV$ has 
been adjusted to zero, it can be seen that to obtain an isolated wall 
configuration the relevant constant of integration must also be taken to 
be zero,
{\be \PP_{\!\perp}=0\, .\fe} 
According to (\ref{Uprime}) this simply gives 
{\be \UU^\pprime=\sqrt{2\VV}\, ,\label{driv}\fe}
and allows one to take the limit $\elle\rightarrow\infty$ in 
(\ref{equmv}) to obtain an expression of the simple form
{\be \UU\{\infty\}=\int \sqrt{2 \VV}\, {\rm d}\llam\, ,\label{Uinf}\fe}
for the surface energy to be minimised in order to obtain the
equilibrium configuration of the isolated wall. 

It can be seen from the formula (\ref{driv}) that this required surface 
energy density function $U$ is interpretable as a generalisation to the 
multiscalar case of what has been referred to in the context of a 
singlet field \cite{ACHL03} as a superpotential, and that it is specifiable 
as the measure of the relevant distance in the field manifold, as evaluated,
not with respect to the original kinetic metric $G_{ij}$, but with respect 
to a conformally modified field energy metric that is given by
{\be {\rm d}\UU^2=\tilde \GG_{ij}\,{\rm d}\Phhi^i\,{\rm d}\Phhi^j
\, ,\label{enmeas}\fe}
with
{\be \tilde \GG_{ij}=2 \VV \GG_{ij}\, .\label{fen} \fe}
The geodesic distance (between the relevant vacua) obtained by minimising 
the integral $\UU\{\infty\}$ given by (\ref{Uinf}) will thus be directly 
identifiable with the required wall tension $\TT$ as given by (\ref{en}). 
The corresponding distribution of the fields as a function of the 
orthogonal distance $x$ is given -- according to (\ref{rest}) -- by a 
metric that is  related to the kinetic metric by a conformal factor that 
is exactly the inverse of the one that gives the energy metric:
{\be {\rm d}x^2=\frac{{\rm d}\llam^2}{2\VV}=\frac{{\rm d}\UU^2}
{4\VV^2}\label{spacm}\, .\fe}

\bigskip\noindent
{\bf 5. Analytically integrable cases of pentavac and monovac doublet 
models}
\medskip

As an illustration of the kind of model that can provide a lattice 
with stable X-type junctions, I proposed as an interesting prototype 
example the particular case of what may be concisely referred to as the 
pentavac model \cite{C04}. This is a model with five equivalent vacuum 
states in a toroidal field space, in which the relevant field variables 
are a pair of phase variables $\phhi$ and $\pssi$ with period $2\pi$ 
with flat kinetic metric given in terms of some fixed mass scale, 
$\eeta$ say, by
{\be {\rm d}\llam^2={\eeta^2}({\rm d}\thheta^2+{\rm d}\chhi^2)=
5\eeta^2({\rm d}\phhi^2+{\rm d}\pssi^2)
\, ,\label{kinec}\fe}
and potential $\VV$ given in terms of some fixed maximum value 
$\VV_{\!\star}$ by
{\be \VV=\frac{\VV_{\!\star}}{4}(
{\rm cos}\,\thheta+{\rm cos}\,\chhi+2)\, .\label{torpot}\fe}
in which
{\be \thheta=2\phhi+\pssi\, , \hskip 1 cm \chhi=2\pssi-\phhi\, .
\label{theph}\fe}

A model of this topologically non-trivial kind is obtainable \cite{C04} 
as a low energy limit $\varepsilon\rightarrow 0$ from a topologically 
simple extended model of with broken U(1)$\times$ U(1) symmetry, of the 
kind is decribed in the appendix.

According to (\ref{fen}) the energy metric obtained from (\ref{kinec}) 
and (\ref{torpot}) on the toroidal space of phase variables $\phhi$ 
and $\pssi$ will be given by the formula
{\be {\rm d}\UU^2={2\eeta^2 \VV}({\rm d}\thheta^2+{\rm d}\chhi^2)
\, ,\label{enmet}\fe}
in which it is useful to  rewrite the formula (\ref{torpot}) for the 
relevant potential as
{\be \VV=\VV_{\!\star}\left(\frac{_1}{^2} {\rm cos}^2\{\thheta/2\}+
\frac{_1}{^2}{\rm cos}^2\{\chhi/2\}\right)\, .\label{Vdef}\fe}
For such a potential, the corresponding geodesic Hamilton Jacobi 
equation, namely
{\be \frac{1}{2\eeta^2 \VV}\left(\left(\frac{\partial \SS}
{\partial\thheta}\right)^2+\left(\frac{\partial \SS}{\partial\chhi}
\right)^2\right)=1\, ,\label{HJ}\fe} 
turns out to have the convenient property of being separable with 
respect to the variables $\thheta$ and $\chi\, .$ By taking the 
Jacobi action variable $\SS$ to be a sum,
{\be \SS=\SS_\thheta+\SS_\chhi\, ,\fe}
of single variable functions, the equation (\ref{HJ}) can be seen
to be reducible to the form
{\be \left(\frac{{\rm d} \SS_\thheta}{{\rm d}\thheta}\right)^2-\eeta^2
\VV_{\!\star}\,{\rm cos}^2\{\thheta/2\}=\eeta^2\VV_{\!\star}\, 
{\rm cos}^2\{\chhi/2\}- \left(\frac{{\rm d} \SS_\chhi}{{\rm d}\chhi}
\right)^2\, ,\fe}
in which the terms on the left depend only on $\thheta$ while
those on the right depend only on $\chhi$, which means that
both sides must be equal to a constant of integration, $\KK$ say.
This means that the corresponding generalised momentum variables,
which are specifiable  according to (\ref{enmet}) as
{\be \calP_\theta={2\eeta^2 \VV}\frac{{\rm d}\thheta}{{\rm d}U}
\, ,\hskip 1 cm \calP_\chhi={2\eeta^2\VV}\frac{{\rm d}\chhi}{{\rm d}\UU}
\, ,\fe}
will satisfy equations of the form
{\be \calP_\thheta^2=\eeta^2 \VV_{\!\star}\,{\rm cos}^2\{\thheta/2\}
+\KK\, ,\hskip 1 cm \calP_\chhi^2=\eeta^2 \VV_{\!\star}\,
{\rm cos}^2\{\chhi/2\}-\KK\, .\label{momv}\fe} 

It is to be recalled that the only trajectories to which this derivation
is applicable are those that begin and end on vacuum states, namely the
field values for which ${\rm cos}^2\{\thheta/2\}={\rm cos}^2\{\chhi/2\}=0
\, .$ In order for the momentum values given by (\ref{momv}) to remain 
real at these end points it is evident that neither $\KK$ nor $-\KK$ can 
be negative, which is only possible if the constant itself vanishes,
{\be \KK=0 \, .\fe}
It can thereby be concluded that the equations of motion for the 
relevant geodesics are given simply by
{\be 4\eeta^2 \VV^2\left(\frac{{\rm d}\thheta}{{\rm d}\UU}\right)^2
=\VV_{\!\star}\, {\rm cos}^2\{\thheta/2\}\, ,\hskip 1 cm
4\eeta^2 \VV^2\left(\frac{{\rm d}\chhi}{{\rm d}\UU}\right)^2
=\VV_{\!\star}\, {\rm cos}^2\{\chhi/2\}\, .\fe}

It can be seen from this that, independently of parametrisation,
the trajectory in phase space will be obtainable by integrating
{\be \frac {{\rm d}\thheta}{{\rm cos}\{\thheta/2\}}=\pm
\frac {{\rm d}\chhi}{{\rm cos}\{\chhi/2\}}\, ,\label{traj} \fe}
and that the corresponding expression for
the  energy variation (in the  ``right'' direction) will
be given by 
{\be {\rm d}\UU=2\eeta\,\sqrt{\VV_{\!\star}}\Big(
{\rm cos}\{\thheta/2\}{\rm d}\thheta\pm{\rm cos}\{\chhi/2\}
{\rm d}\chhi\Big)\label{trajen}\, .\fe}

An example of a simple wall trajectory (such as the straight line 
AB in Fig. \ref{ABCD}) is obtainable by holding $\chhi$ fixed at a 
value such that ${\rm cos}\,\{\chhi/2\}=0$ (so that the ratio on the 
right of (\ref{traj}) is indeterminate) and letting $\thheta$ vary from
$-\pi$ to $\pi$. According to (\ref{spacm}) the wall profile of the 
field $\theta$ as a function of the orthogonal space coordinate $x$ 
will be given by the relation
{\be {\rm cot}\left\{\frac{\thheta-\pi}{2}\right\}={\rm sinh}\left
\{\frac{x}{\delt_\star}\right\}\, .\fe}
in which the effective wall thickness scale $\delt_\star$ is 
determined by the relevant mass scale $\mm_\star$ as
{\be \delt_\star=\frac{2\eeta}{\mm_\star^{\,2}}\, ,
\hskip 1 cm \mm_\star^4=\VV_{\!\star} \, .\label{delta}\fe}
In this case there will be no contribution from the right of 
(\ref{traj}), so the complete integral $\UU\{\infty\}$ giving the 
wall tension $\TT_{\rm I}$ works out as
{\be \TT_{\rm I}=4\,\mm_\star^{\,2} \eeta=2\,\VV_{\!\star}\,
\delt_\star\, .\label{simten}\fe}

To get a compound wall trajectory (of the kind illustrated by the 
curves AC in Fig. \ref{ABCD}) the variable $\chhi$ is no longer 
held fixed but is also allowed, like $\thheta$ to vary from $-\pi$ 
to $\pi$. Such a trajectory is obtainable by choosing $\pm=+$ in 
equation (\ref{traj}),  which is then easily integrable to give
{\be {\rm cot}\left\{\frac{\chhi-\pi}{4}\right\}=\kappa\,
{\rm cot}\left\{\frac{\thheta-\pi}{4}\right\}\, ,\label{kappa}\fe}
where $\kappa$ is a dimensionless constant of integration. It takes the 
value $\kappa=1$ in the trivial case of the straight diagonal trajectory 
given by the equation $\chhi=\thheta$. For any non-zero value of 
$\kappa$, the separate energy contributions from the pair of terms 
on the right of (\ref{trajen}) will integrate to the same final 
result. Thus the various trajectories given by different values of 
$\kappa$ in (\ref{kappa}) are energetically degenerate, all giving 
the same compound wall energy
{\be \TT_{\rm I\!I}=4\VV_{\!\star}\,\delt_\star\, .\fe}
The condition of being just twice the simple wall energy is an obvious 
consequence of the fact that although their topology is globally 
interwoven, the separate field combinations $\thheta$ and $\chhi$ 
behave locally as a pair of decoupled scalars, so their wall 
configurations can travel through each other without interaction. 

According to the analysis in the preceding sections, this feature
{\be \TT_{\rm I\!I}/\TT_{\rm I}=2\, ,\fe}
is just what is marginally needed to ensure stability (against 
disintegration into pairs of Y junctions) of the X-junctions,  
whatever their crossing angle $\alpha$ may be (not just when it is near 
a right angle as would be needed in models providing a lower value of 
$\TT_{\rm I\!I}/\TT_{\rm I}\, .$) Without going into these quantitative 
details, the stability, as previously claimed \cite{C04}, of the 
X-junctions in this  particular case was heuristically  obvious in 
advance from the lack of any mechanism of disintegration (into pairs 
of Y junctions) in view of the  effective absence of interaction 
between the separate $\thheta$ and $\chhi$ fields that respectively 
characterise the intersecting walls. (For sufficiently small but non-zero 
values of the parameter $\varepsilon$ in the corresponding topologically 
simple extended model \cite{C04} described in the appendix, continuity 
implies that X-junction stability will still hold except for 
correspondingly small -- meaning highly acute --  values of the 
intersection angle $\alpha$.)

Although there is no interaction between walls produced by the separate
variation of $\thheta$ and $\chhi$, it is important to notice that there 
will be an interaction between the parallel walls in the periodic 
solution produced by the variation of a single one of the separate 
variables, $\thheta$ say, for a finite value of the wavelength $\elle$ as 
given by the formula (\ref{length}) for the separation distance. In such 
a lasagne type configuration there will be an interaction, interpretable
as an effect of mutual repulsion, due to the build up, as $\elle$ 
decreases, of the corresponding orthogonal pressure $P_{\!\perp}$. In the 
separable case under consideration, it can be seen from (\ref{length})
and (\ref{equmv}) that, in terms of the ``first'' and ``second'' kinds 
of elliptic integral \cite{AS},
{\be K\{\mmu\}=\int_0^1\frac{{\rm d}t}{\sqrt{(1-t^2)(1-\mmu\, t^2)}}
\, ,\hskip 0.6 cm  E\{\mmu\}=\int_0^1\sqrt\frac{1-\mmu\,t^2}{1-t^2}\,
{\rm d}t\, ,\fe}
the relations between  pressure, separation, and the enthalpy density
$H$ will be given by
{\be \frac{\elle}{2\delt_\star}=\sqrt{\mmu}\,K\{\mmu\}\, ,\hskip 1 cm
\frac{\HH}{\TT_{\rm I}}=\frac{E\{\mmu\}}{\sqrt{\mmu}}\, ,\hskip 1 cm 
\mmu=\frac{\VV_{\!\star}}{\VV_{\!\star}+2\PP_{\!\perp}}
\, ,\label{implicit}\fe}
where $\delt_\star$ is the effective wall thickness as given
by (\ref{delta}). 

Since the asymptotic behaviour of the elliptic integral is known 
\cite{AS} to be characterised by the condition that
$ K\{\mmu\}+{\rm ln}\{\sqrt{1-\mmu}/4\}\rightarrow 0$ as 
$\mmu\rightarrow 1 ,$ it can be seen to follow that in the long 
wavelength limit $\elle\rightarrow\infty$, the relevant pressure will 
be given asymptotically by 
{\be \PP_{\!\perp}\sim 8 \VV_{\!\star} \, {\rm e}^{-\elle/\delt_\star}
\label{Pell}\, ,\fe}
while in terms of the tension $\TT_{\rm I}$ of an isolated wall, as 
given by (\ref{simten}), it can be seen from (\ref{pdef}) that the 
surface energy density per period will be given by the asymptotic 
formula
{\be \UU\simeq \TT_{\rm I}\left(1+4 {\rm e}^{-\elle/\delt_\star}\right)
\, . \label{Uell}\fe}
According to (\ref{vel}) the velocity of large lengthscale longitudinal 
perturbations will be given, in this long wavelength limit 
$\elle\gg\delt_\star$, by an expression of the corresponding form
{\be \vv_\perp\sim \frac{2\elle}{\delt_{\!\star}}\,\,
 {\rm e}^{-\elle/2\delt_\star}\, ,\label{vper}\fe}
whose reality is what guarantees stability against bunching.

\bigskip\noindent
{\bf 6. Beta monovac and multivac winding models}
\medskip

The fields $\phhi$ and $\pssi$ in the pentavac model on the preceeding 
section were set up as the respective phases of unimodular complex 
field variables ${\rm e}^{i\phhi}$ and ${\rm e}^{i\pssi}$ that could 
themselves be considered to have been obtained by imposition of a low
energy restraint on corresponding complex variables $\Phhi$ and $\Pssi$ 
whose modulus at much higher energy would no longer be restrained.
A new model that would be physically indistinguishable in in the low 
energy limit characterised by the unimodularity condition could be set 
up by instead taking the independent fields to be the unimodular complex 
variables ${\rm e}^{i\thheta}$ and ${\rm e}^{i\chhi}$ that are defined 
by the combinations $\thheta$ and $\chhi$ specified by (\ref{theph})

Whereas the toroidal space occupied by the original pair of unimodular 
field variables ${\rm e}^{i\phhi}$ and ${\rm e}^{i\pssi}$ can be 
covered by any one of the four large square patches bounded by dotted 
lines in Fig. \ref{ABCD}, the (much smaller) toroidal space occupied by 
the new pair of unimodular field variables ${\rm e}^{i\thheta}$ and 
${\rm e}^{i\chhi}$ can be covered just by the small square with vertices 
at the positions marked A,B,C,D, which in the new model are to be 
identified, so that there will now be just a single vacuum state instead 
of five. In the new ``monovac'' model (a doublet generalisation of the 
singlet sine Gordon equation) that is obtained in this way, a simple 
wall of the kind exemplified by the trajectory AB will still be a 
topologically stable membrane defect, but of the kind known 
specifically as a ``winding'', to distinguish it from the more commonly 
discussed  kind of ``open kink'' for which (as in the pentavac case) 
the vacuum states on either side are of distinct varieties.

Both the original pentavac model and the new monovac model can be 
regarded as special separable cases within larger families of 
respectively pentavac and monovac models for which the potential
is given in terms of a fixed index $\bbeta>0$  by an expression 
of the form
{\be \VV=\frac{\mm_\star^{\,4}}{2}\left( {\rm cos}^2\{\thheta/2\}+
{\rm cos}^2\{\chhi/2\}\right)^\bbeta\, ,\label{beta}\fe}
which includes the special separable example (\ref{Vdef}) as the 
particular case for which $\bbeta=1$.

It is apparent that -- for sufficiently large values of the angle 
$\delta$ of deviation from orthoginality in Fig. \ref{XtoY} -- the 
possibility of instability of an X type junction with respect to 
disintegration into Y type junctions \cite{AMMM06} will occur in such 
(monovac or pentavac) models if and only if the strict inequality 
$\bbeta<1$ is satisfied. By comparing the double wall trajectory
parametrised by the symmetric relation $\chhi=\thheta$ with the simple 
wall  trajectory given by the fixed value $\chhi=\pi$ it can be seen 
from (\ref{beta}) that the energy $\TT_{\rm I\!I}$ of the former (as 
represented by the diagonal $AC$ in Fig. \ref{ABCD}) will be related to 
the energy $\TT_{\rm I}$ of the latter (as represented by AB) by
{\be \TT_{\rm I\!I}=(\sqrt 2)^{1+\bbeta} \,\TT_{\rm I}
\, . \label{trat}\fe}
The actual value of the simple wall tension in such a case can be seen
to be given in terms of the thickness parameter $\delt_\star$ defined
by (\ref{delta}), and an order of unity factor $I_\bbeta ,$ by
{\be \TT_{\rm I}=4\, \mm_\star^{\,2}\,\eeta\,
I_\bbeta\, ,\hskip 1 cm I_\bbeta=\int_0^1\frac
{y^\bbeta\,{\rm d} y}{\sqrt{1-y^2}}\, , \label{tenfo}\fe}
which in the particularly interesting case $\bbeta=2$ gives
$ \TT_{\rm I}= \pi\,\mm_\star^{\,2}\eeta .$

It evidently follows from (\ref{trat}) that for $\bbeta>1$ the double 
wall configuration (such as  AC) will be unstable with respect to 
decomposition into a pair of simple walls (such as AB and BC) between 
which there will be an effective repulsion. On the other hand, for 
$\bbeta<1$ a complementary pair of simple walls (such as AB and BC) 
would be attractive, and in such a case, according to (\ref{crita}) an 
X junction will be unstable if but only if the deviation angle $\ddelta$ 
in (\ref{delt}) is large enough to satisfy the strict inequality
{\be {\rm sin}\,\ddelta > 2^\bbeta -1 \, ,\fe}
a condition that would evidently be impossible for  $\bbeta\geq 1$. 

The foregoing considerations remain true when -- to be cosmologically
realistic -- the monovac doublet model is extended to a monovac triplet
model by the inclusion of a third unimodular scalar field,
${\rm e}^{i\ssigma}$ say, acting in the same way as the others, so 
that the complete kinetic metric will be given in terms of a 
relatively large mass scale $\eeta$ 
by {\be {\rm d}\llam^2={\eeta^2}({\rm d}\thheta^2+
{\rm d}\chhi^2+{\rm d}\ssigma^2)\, ,\label{kinet}\fe}
while in terms of a smaller mass scale $\mm_\star$ the complete 
potential function will take the form
{\be \VV=\frac{\mm_\star^{\,4}}{2}\left( {\rm cos}^2\{\thheta/2\}+
{\rm cos}^2\{\chhi/2\}+{\rm cos}^2\{\ssigma/2\}\right)^\beta
\, ,\label{betam}\fe}
which reduces to (\ref{beta}) wherever ${\rm cos}\{\ssigma/2\}=0$.

It is to be remarked that the potential functions of the same form 
can be used to characterise triplet models with not just one but many 
energetically degenerate vacuum states, by supposing that -- instead 
of being obtained directly as phases of unimodular complex numbers -- 
the variables $\thheta\, ,\, \chhi\, ,\, \ssigma\, ,\,$ are obtained 
as linear combinations of quantities $\phhi\, ,\,\pssi\, ,\, \xsi$ 
that actually do have this property, meaning that they are phases of 
corresponding field variables ${\rm e}^{i\phhi}\, ,\, {\rm e}^{i\pssi}
\, ,\, {\rm e}^{i\xsi}\, .$  The most obvious possibility is to halve 
the periodicity in each of the three principle directions so as to 
obtain an octovac model -- meaning one with $2^3\,(=8)$ vacuum states 
-- by setting
{\be \thheta=2\phhi\hskip 1 cm\chhi=2\pssi\, ,\hskip 1cm
\ssigma=2\xsi\, ,\label{doub}\fe}
which entails that the kinetic metric (\ref{kinet}) will be 
expressible as
 {\be {\rm d}\llam^2=4{\eeta^2}({\rm d}\phhi^2+
{\rm d}\pssi^2+{\rm d}\xsi^2)\, .\label{kine}\fe}
In the same way, dividing the periods by three instead of two, one can 
obtain a triplet model with $3^3\,(=27)$ vacuum states by setting
{\be \thheta=3\phhi\hskip 1 cm\chhi=3\pssi\, ,\hskip 1cm
\ssigma=3\xsi\, ,\label{trip}\fe}
for which the kinetic metric (\ref{kinet}) will be expressible as
 {\be {\rm d}\llam^2=9{\eeta^2}({\rm d}\phhi^2+
{\rm d}\pssi^2+{\rm d}\xsi^2)\, .\label{kin}\fe}
An alternative 27-vac triplet model with the same kinetic metric
(\ref{kin}) is obtainable -- as a triplet analogue of the pentavac 
doublet model (\ref{theph}) discussed above -- by replacing 
(\ref{trip}) by a ``tilted'' (but still conformal) transformation 
relation of the form
{\be \thheta=\phhi-2\pssi+2\xsi\, ,\hskip 1 cm
\chhi=2\phhi-\pssi-2\xsi\, ,\hskip 1cm
\ssigma=2\phhi+2\pssi+\xsi\, .\label{tilt}\fe}
The property of providing same number of distinct vacuum states
as the more trivial model characterised by the simple transformation
(\ref{trip}) follows from the easily verified condition that the 
Jacobean determinant of the tilted transformation (\ref{tilt}) has the 
same value, namely 27.

By a mutually orthogonal superposition of the lasagne type configuations
corresponding to variation of $\thheta$, $\chhi$ and $\ssigma$ 
respectively, it can be seen that for $\bbeta>1$ such models can provide 
a robustly stable cubical (rather than merely square) lattice of the kind 
needed for the averaged stress tensor to be isotropic. For a single one
of these lasagne configurations, corresponding just to the variation
of $\thheta$ say, with the other two variables fixed at the energy
minimising values ${\rm cos}^2\{\chhi/2\}={\rm cos}^2\{\ssigma/2\}=0$,
it can be seen from (\ref{length}) that for $\bbeta>1$ -- instead of 
falling of in the exponential manner described by (\ref{Pell}) -- the 
dependence on the separation distance $\elle$ of the orthogonal 
pressure will be given asymptotically, as $\elle\rightarrow\infty ,$ 
by a power law of the form
{\be \PP_{\!\perp}\sim \frac{\mm_\star^{\,4}}{2}\left(\frac
{\delt_{\bbeta}}{\elle}\right)^{2\bbeta/(\bbeta-1)} \, ,\label{asp}\fe}
in which $\delt_\beta$ is of the same order of magnitude as the 
original thickness scale $\delt_{\!\ast}$ defined by (\ref{delta}), 
to which it is related by the specification
{\be \delt_\bbeta=J_\bbeta \,\delt_\star\, ,\hskip 1 cm
J_\bbeta=\int_0^\infty\frac{{\rm d} z}{\sqrt{z^{2\bbeta}+1}}\, ,\fe}
which, in the quartic case $\bbeta=2$ provides the expression 
$J_2=4\,\Gamma^2\{5/4\}/\sqrt\pi=\varsigma\pi/2$ in terms of a numerical 
factor close to unity given by $\varsigma\simeq 1.18$. 

It follows that the analogue for $\bbeta>1$ of the asymptotic formula 
(\ref{Uell}) for the surface energy density of each slab of thickness
 $\elle$ will be given by
{\be \UU\simeq \TT_{\rm I}+\frac{\bbeta-1}{\bbeta+1}\elle 
\PP_{\!\perp}\, ,\fe}
and that according to (\ref{vel}) the analogue of (\ref{vper}) for
the squared velocity of large lengthscale longitudinal perturbations 
takes the form
{\be \vv_\perp^2\sim \frac{\bbeta\,I_\bbeta}{2(\bbeta -1) J_\bbeta}
\left(\frac{\delt_\bbeta}{\elle}\right)^{(\bbeta+1)/(\bbeta-1)}\, ,\fe}
which is manifestly positive, as required for stability, whenever
$\bbeta>1 .$

\bigskip\noindent
{\bf 7. Macroscopic comportment}
\medskip

One of the objections raised against the idea of attributing the
cosmological acceleration to a regular domain wall lattice was the 
lack of a plausible mechanism to prevent parallel walls from
drifting together in the long run and eventually undergoing mutual
annihilation. It is therefore to be emphasised that this is not a 
danger in the kind of monovac model advocated here, in which the 
parallel wall number density is interpretable as a conserved 
topological winding number, and the parallel walls are protected 
against getting too close by the effect whereby, when the separation 
distance $\elle$ gets too small, the orthogonal pressure $\PP_{\!\perp}$ 
in (\ref{length}) will cease to be negligibly small, and will build 
up so as to produce a mutually repusive force.

The simplest cosmologically viable category is that of the separable 
-- $\bbeta=1$ -- kind of monovac triplet model (effectively a 
superposition of three independent sine-Gordon singlets) which 
provides a cubic lattice with a cosmologically comoving lengthscale 
$\elle$ and hence cell number density $\nn=1/\elle^3$ for which the 
relations (\ref{implicit}) implicitly provide a corresponding equation 
of state whereby the cosmological energy density $\rrho\, ,$ will be 
given by
{\be \rrho=3\frac{\UU}{\elle}\, , \hskip 1 cm \UU=\HH-\PP_{\!\perp} \elle 
\, ,\label{wallden}\fe}
while the corresponding isotropic 
cosmological pressure $\PP\, ,$ which will be given simply by
{\be \PP=\PP_{\!\perp}-2\frac{\UU}{\elle}=3\PP_{\!\perp}
-2\frac{\HH}{\elle} \, ,\fe}
which means that the cosmologically important ratio $\ww=\PP/\rrho$ will 
be given by the formula
{\be \ww=-\frac{2}{3}+\frac {\PP_{\!\perp}}{\rrho}\, .\label{www}\fe}

At a much earlier epoch, when the winding wavelength $\elle$ would have
been comparable with or even much less than the wall thickness scale
$\delt_\star$ given by (\ref{delta}), it can be seen from (\ref{hype}) 
that one would have had 
{\be \ww\simeq -\frac{1}{3} \hskip 1 cm {\rm with} \hskip 1 cm
\frac{\PP_{\!\perp}}{\rrho}\sim\frac{1}{3}\, ,\hskip 1 cm
\rrho\sim \frac{6\pi^2\eeta^2}{\elle^2}\, .\fe}
However at the present epoch it is to be presumed that the 
(cosmologically comoving) lengthscale $\elle$ will have become very 
much larger, $\elle\gg\delt_\star , $ and hence  that it will be 
possible to neglect the final term in (\ref{www}), which would fall off 
rapidly with a negative power law dependence on $\elle$ for $\bbeta>1 ,$ 
and with an exponential dependence given by the asymptotic formula
{\be \frac {\PP_{\!\perp}}{\rrho}\sim \frac{4\,\delt_\star}{3\,\elle}\,
{\rm e}^{-\elle/\delt_\star}\, ,\fe}
for the case $\bbeta=1$, so that (in all these cases) one would ultimately 
attain the observationally admissible \cite{Silk04} value given,
using (\ref{tenfo}), by
{\be \ww\simeq -\frac{2}{3} \, ,\hskip 1 cm {\rm with}\hskip 1 cm
\rrho\approx 10 \frac{\mm_\star^{\,2}\eeta}{\elle}\, .\label{denl} \fe}

For a simple perfect fluid such an equation of state would of course be 
unacceptably unstable. The mechanism for the stabilisation confirmed in 
the present work is of exactly the kind that is describable by treating 
the large scale averaged system as a perfectly elastic solid (in which 
the three relevant fields $\thheta$, $\chhi$, $\ssigma$ would be comoving 
``base space'' coordinates) of the kind originally envisaged by Bucher 
and Spergel\cite{BS98}, of which it is thus a prefect example. 

In order to have become dominant, the density (\ref{denl}) must have 
recently reached the order of magnitude of the baryonic mass density 
which is given in terms of the proton mass $\mm_{\rm p}$ and the 
cosmological temperature $\Thheta$ by $\rrho_{\rm b}\approx 
10^{-8}\mm_{\rm p}\,\Thheta^3 . $ Since the temperature has a 
contemporary value given in terms of the electron mass $\mm_{\rm e}$ by
$\Thheta_{\rm c}\approx 10^{-9}\mm_{\rm e}  ,$ it can be seen to follow 
that the ratio of the contempory value $\elle_{\rm c}$  of the mesh 
spacing $\elle$ to the value (in the millimeter range) of the contempory 
thermal wavelength $\approx  \Thheta_{\rm c}^{ -1}$ must be given by
{\be \Thheta_{\rm c}\elle_{\rm c}\approx 10^{27}\frac{\eeta}{\mm_{\rm p}}
\left(\frac{\mm_\star}{\mm_{\rm e}}\right)^2\, . \label{domcon}\fe}

According to the line of reasonning in the preceding analysis \cite{C04}, 
the Kibble type wall formation mechanism that was envisaged would
be likely to provide values of this length ratio $\Thheta_{\rm c}
\elle_{\rm }$ of the order of  $10^{14}$, which would be obtained for 
$\mm_\star \approx 10^{-5}\eeta $ with $\eeta\approx \mm_{\rm e} ,$ or
for $\mm_\star\approx 10^{-2}\eeta$ with  
$\eeta\approx 10^{-2}\mm_{\rm e} .$ These values are 
comparatively small (not comparable, as was stated due to a transcription 
error) with respect to the interstellar distance scale, of the order of 
several parsecs, which is what would be obtained for $\elle_{\rm c}$ if 
one used the rather larger mass values $\mm_\star\approx \eeta
\approx 10^{-1}\mm_{\rm e}$ that were suggested by an analysis of the 
cruder kind used in earlier work \cite{BBS99}. However it now seems that 
all such reasoning, whether in its earlier form or in the more refined 
\cite{C04} version, should be considered to be effectively obsolete,
because its generic outcome will be the formation of disorganised 
lattice configurations that would  be insufficiently stable, unlike the
 superposed lasagne type configurations envisaged here.

The whole question of the wall formation process needs to be entirely
reconsidered in the context of the stabilisation mechanism considered
here, which does not work for random wall lattice configurations,
but is essentially dependent on the existence of effectively conserved
winding numbers with sufficiently large values, representing the effect 
of systematic winding in the same sense over a cosmologically large scale. 
Even for an underlying field model of the appropriate kind (as 
illustrated by the examples considered above) the ordinary kind of 
Kibble mechanism considered so far \cite{BBS99,C04} would give rise to 
random combinations of positive an negative windings, corresponding to 
what might be called walls and anti-walls, which in the long run would 
be unstable with respect to mutual annihilation. The work in the 
following section suggests, however,  that it be possible to overcome 
this problem by incorporating the effect of the kind inflationary process 
\cite{Guth81} that has been developped for dealing with what is known as 
the horizon problem. On the basis of this revised analysis it will be 
found that the most plausible mean mass values are not smaller but larger 
than those that were originally envisaged, with the implication that the 
present mesh lengthscale $\elle_{\rm c}$ should actually be expected to be 
of intergalactic order.

\bigskip\noindent
{\bf 8. An inflationary mechanism?}
\medskip

The horizon problem was posed in the context of traditional cosmology
by the need to account for the approximate homogeneity -- indicative
of causal contact in the past -- that is observed all the way out to the 
present value of the Hubble radius $\Rh$ as defined, in terms of the 
proper time derivative $\dot \aa$ of a comoving lengthscale $\aa$ say, 
by $1/\Rh=\dot \aa/\aa .$ This suggests that the  
radius, $\xxi_{\rm C}$ say, of effective causal correlation must be 
substantially larger than this, $\xxi_{\rm C}\gg \Rh$. The problem is 
that in the traditional scenario of a decelerating radiation dominated 
universe this Hubble radius has the same order of magnitude as this 
``particle horizon''  radius  $\xxi_{\rm C} .$ To get over this, the 
idea put forward by Guth \cite{Guth81} was that such a restriction would 
no longer apply if, at some stage in the past, there had been a 
sufficiently long period of inflation during which the universe was 
subject to acceleration, $\ddot\aa>0$ of the kind that seems to have 
recently started again.

When such considerations were originally introduced by Guth \cite{Guth81},
the intended application was to the breakdown of grand unification
leaving a relic distribution of monopole particles whose density per 
Hubble volume was assumed to have been at least of the order of unity at 
the  relevant initial time $\tt_{\rm i} ,$  implying an initial number 
density $\nn_{\rm i}\gta \xxi_{\rm Ci}^{-3}.$ The monopole problem was 
actually even worse than commonly supposed since such a lower limit 
can plausibly be strengthenned by taking account of the consideration 
that thermal fluctuations corresponding to the temperature 
$\Thheta_{\rm i}$ (in Planck units) at that epoch might have initially 
produced roughly of the order of one of the relevant particles or 
antiparticles per thermal volume $\Thheta^{-3}$, of which there would 
have then have been $N\approx(\xxi_{\rm Ci}\Thheta_{\rm i})^3$  in the 
correlated volume under consideration. As the causal correlation radius
increased, the created number in the corresponding volume would also
have increased by a factor $(\xxi_{\rm C}/\xxi_{\rm Ci})^3
(\aa/\aa_{\rm i})^{-3}$ to give
{\be N\approx (\xxi_{\rm C}\Thheta_{\rm i}\aa_{\rm i}/\aa)^3\, .\fe}
Due to their causal interaction, these particles and antiparticles would 
however have undergone mutual annihilation, except for a relatively small 
excess with the same sign (e.g. with no  antiparticles) whose number 
would have had a root mean square value that can be estimated by a random 
walk argument to be of the order of $\sqrt{N} ,$ as given by the estimate
{\be \sqrt{N}\approx (\xxi_{\rm C}\Thheta/\ZZ_{\rm i})^{3/2}\, .\fe}
in which the thermal evolution formula
{\be \Thheta/\Thheta_{\rm i}=\ZZ_{\rm i}\, \aa_{\rm i}/\aa \, ,\fe} 
has been used to define a reheating factor $\ZZ_{\rm i}$ factor of the 
kind that was supposed by Guth to have become extremely large, and that 
for consistency with the second law of thermodynamics must at least 
increase monotonically from an initial value $\ZZ_{\rm i\, i}=1 .$ 
In terms of the amplification factor
{\be \aaleph=\xxi_{\rm C}/\Rh\, ,\fe}
this means that the order of magnitude of the number density with which 
they would have ultimately emerged would be expressible as
{\be \nn \approx \left(\frac{\Thheta}{\aaleph\ZZ_{\rm i}  \Rh}
\right)^{3/2}\label{ndens}\, .\fe}
Since the ratio of the present cosmological radius $\Rh{_{\rm c}}
\sim 10^{59}$ to the thermal lengthscale given by the present 
cosmological temperature $\Thheta_{\rm c}\approx 10^{-31}$ is given 
roughly by $\Rh{_{\rm c}}\Thheta_{\rm c}\approx 10^{28} ,$ the implication 
is that the number of such particles in a Hubble volume would be given 
now by
{\be \nn_{\rm c} \Rh{_{\rm c}}^{\, 3} \approx \left(\frac{\Rh{_{\rm c}}
\Thheta_{\rm c}}{\aaleph_{\rm c} \ZZ_{\rm i\,c}}\right)^{3/2}
\lta 10^{42} \, , \label{Guth}\fe}
in which the lower limit on the right is obtained by postulating the 
order of unity values that are the smallest conceivable possibilities for 
the inflation amplitude $\aaleph_{\rm c}$ and the reheating factor 
$\ZZ_{\rm i\, c}$ at the present epoch. Guth's idea \cite{Guth81} for the 
latter, in the context of the monopoles predicted by the Grand 
unification theory (which was more fashionable then than now) was to 
postulate an enormous value $\ZZ_{\rm i\,c}\sim 10^{28}$ in order to get 
the right hand side of (\ref{Guth}) down to near the order of unity. 

At the opposite extreme, in the context of the baryon formation problem, 
it is to be noted that although it is rather large, the maximum 
value on the right of (\ref{Guth}) is barely the square root of the 
value -- nearer $10^{76}$ -- of Dirac's cosmological baryon number, thus
falling far short of what would be needed even to make a single star, so 
it can not be hoped that a mechanism of this kind would suffice to solve 
the problem of the excess of baryons over antibaryons. 

What such a mechanism might conceivably do, however, would be to solve the 
analogous problem of getting the excess of positive over negative windings 
that would be needed to provide the kind of domain wall lattice discussed 
above, in which the lattice spacing $\ell$ would correspond to a cell 
number density $\nn=\elle^{-3} ,$ so that the formula (\ref{ndens}) would 
provide for the the relevant contempory values the analogous estimate
{\be \frac{\Thheta_{\rm c}\elle_{\rm c}}{\sqrt{\aaleph_{\rm c}\ZZ_{\rm i\,c} 
}}\approx\sqrt{\Rh{_{\rm c}}\Thheta_{\rm c}}\approx 10^{14}\, ,\label{AZ}\fe}
which means that, even for very moderate values of the ratios 
$\aaleph_{\rm c}$ and $\ZZ_{\rm i\,c} ,$ the contempory value $\elle_{\rm c}$
of the wall spacing now would at least be large compared with the size of 
the solar system. To avoid having $\elle_{\rm c}\gta \Rh{_{\rm c}} ,$ which 
would mean that the walls would be entirely inflated away (in the manner 
envisaged by Guth for the monopoles of grand unifcation) it can be
seen that the inflation factor must satisfy 
{\be  \sqrt{\aaleph_{\rm c}\ZZ_{\rm i\,c} }\ll 10^{14}\,  .\label{Zlim}\fe} 

Comparing (\ref{AZ}) with the matching condition (\ref{domcon}) one sees 
that it requires that the relevant mass ratios should to be related by
{\be \frac{\eeta}{\mm_{\rm e}}\left(\frac{\mm_\star}{\mm_{\rm e}}\right)^2
\approx 10^{-8} \sqrt{\aaleph_{\rm c}\ZZ_{\rm i\, c}  }\, . \fe}
On the assumption  that the potential mass scale is small compared with the 
kinetic mass scale, meaning $\mm_\star\ll \eeta ,$ it can thereby be see 
to follow from (\ref{Zlim}) that the lighter mass scale must be subject
to a rather severe upper mass limit given by
{\be \mm_\star\ll 10^2 \mm_{\rm e} \, .\label{low}\fe}

An opposing restriction comes from the astrophysical consideration that,
to avoid undermining the experimentally well confirmed scenario
of chemical element formation in the temperature regime
$\Thheta\lta \mm_{\rm e} ,$ the postulated inflation following the
relevant phase transition,  with $\Thheta_{\rm i}\lta\eeta ,$ must have 
been finished before this stage, which evidently imposes the requirement 
that the heavier mass scale should be subject to a lower mass limit 
given by
{\be \eeta\gg \mm_{\rm e} \, .\label{hi}\fe}

\medskip\noindent
{\bf 9. Conclusions}
\medskip

To compute the stability of an X type junction in the framework
of a particular field theoretical model it is not necessary to
work with the essentially two dimensional geometry that is actually
involved: it will suffice to consider the effectively one dimensional 
problems of plane wall configurations of simple and double type, in 
order to obtain the corresponding energy densities $\TT_{\rm I}$ and 
$\TT_{\rm I\!I}$ whose ratio is needed. A right angle junction 
($\aalpha=\pi/2$ in Fig. \ref{XtoY}) will always be stable so long as 
{\be \TT_{\rm I\!I}/\TT_{\rm I}>\sqrt{2} \, ,\label{cond}\fe}
but a larger ratio will be needed for stablity of X junctions
at other angles.

For models of the forced harmonic kind it has been shown that the 
required tensions are obtainable as the field space distances between 
the relevant vacua, as measured with respect to the energy metric 
given by (\ref{fen}). For the separable models such as the pentavac 
doublet model~\cite{C04} that was subject to question \cite{AMMM06},
this geodesic method has been used for the explicit evaluation of
the relevant tensions, and it has been shown that the stability
condition is indeed satisfied. Finally attention has been drawn to a
cosmologically more promising class of monovac triplet models
for which the stability condition is guaranteed whenever
the relevant index $\beta$ exceeds unity.

At a macroscopic level, the particular models considered here are 
well described by the formalism of an elastic solid medium, but it 
is to be observed that,  convenient though it is \cite{BS98}, this
feature is  not essential for their stability. The winding stabilisation 
mechanism for the superposed lasagne type configurations envisaged 
here would work just as well if, instead of three (the minimum 
compatible with an isotropic total stress tensor) a higher number of 
independent scalar fields were invoked. In that case one would obtain 
more than triply superposed lasagne type configurations which would be 
just as stable, but in which there would be too many degrees of 
freedom for the system to behave as a simple elastic (or even 
hyperelastic \cite{C06}) solid unless coupled  with something else that 
ensured the required cohesion. The lack of a plausible cohesion mechanism
was one of the weak points in the kinds of solid lattice scenario that 
were originally considered. It is therefor to be emphasised that this 
drawback does not apply to the stability mechanism considered here, 
for which no such cohesion mechanism is needed. 

The drawback that remains is, as remarked in the introduction, the 
need for the system to have been created in a configuration that is
wound up coherently in the same sense (without string winding defects,
and with only ``positive'' walls as opposed to ``anti-walls'') over a 
sufficiently large cosmological scale. This last drawback is serious, 
but (particularly in view of the likely involvement of an anthropic 
selection mechanism) it should not be considered to be automatically 
fatal. It should rather be taken 
as a challenge that is comparable with (and perhaps related to) the 
unavoidable challenges posed by the problem of the parity breaking that 
is associated with the observationally well established preponderance 
of ordinary matter over anti-matter \cite{OMA05}, and by the -- rather
 less intractible and perhaps more directly relevant -- horizon
problem, which provides a major incentive for the hypothesis of 
cosmological inflation, and also motivates consideration of effects 
of multiconnectedness \cite{RWULL04}.

The provisional investigation in the preceeding section suggests that it 
may indeed be possible for an appropriate inflation scenario to provide 
what is needed.  An attractive feature of this approach is that while the 
concomitant restrictions (\ref{Zlim}), (\ref{low}), (\ref{hi}) rule out 
more exotic mass values, they can be satisfied in a very reasonable way 
by postulating that the relevant potential and kinetic mass scales have 
the most obvious orders of magnitude, namely  
{\be \mm_\star\approx \mm_{\rm e}\, ,\hskip 1 cm \eeta\approx \mm_{\rm p} 
\, .\fe} 
This would require that the relevant late inflation stage -- initiated
with temperature $\Thheta_{\rm i}$ at or below the B.e.v. level -- should 
have provided a mean inflation factor $\sqrt{\aaleph_{\rm c}\ZZ_{\rm i\, c}} 
\approx 10^{11}, $ in which $\ZZ_{\rm i\,c}$ is the reheating factor 
and $\aaleph_{\rm c}$ is the factor (if any) by which the relevant range 
of causal influence exceeds the Hubble radius $\Rh{_{\rm c}}$. The ensuing 
wall mesh scale would then be 
{\be \elle_{\rm c} \approx 10^{-3} \Rh{_{\rm c}} \, ,\fe} 
which means that, in a double inflation scenario, a previous such mean
inflation factor of $10^3$ (due perhaps just to reheating by $10^6$) would 
have sufficed to sweep away any grand unification monopoles. It is 
noteworthy that, in such a scenario, the contempory  mesh length 
$\ell_{\rm c}$ would be in rather satisfactory agreement with the order of 
the ten megaparsec scale to which (in the context of double inflation) 
attention has been drawn \cite{PPS94} as the intergalactic threshold for 
an enhancement of the power spectrum of the observed structure. This 
agreement would also be compatible with a more extreme mass parameter 
ratio and higher inflation temperature, such as might be given, with 
$\Thheta_{\rm i}$ above the T.e.v. level, by 
$\mm_\star\approx 10^{-2}\,\mm_{\rm e}$ with $\eta
\approx 10^4 \,\mm_{\rm p}$.

\vfill\eject

\bigskip
\noindent
{\bf Acknowledgements}. I wish to thank R. Battye, M. Bucher,
E. Chachoua, and A. Moss for stimulating conversations.

\vfill\eject

\bigskip\noindent
{\bf Appendix A: Derivation from extended field models.}
\medskip

In a universe with appropriate toroidal topology (so as to admit 
configurations without any string defects) the field models 
considered above might be postulated to have an essentially 
fundamental status. However in might be computationally convenient 
(and in other contexts, such as that of Kibble type mechanisms 
\cite{GH03, APV04, OMA05}, more natural) to consider such 
topologically non-trivial field models to be derived as approximate 
effective models from extended field spaces of higher dimension but with 
simple geometry, in which the relevant total potential $\VV_{\rm tot}$ 
contains a sufficiently steeply variable confining contribution, 
$\VV_{\rm cof}$ say, having a degenerate minimum on a submanifold 
characterised by the vanishing of a set of functions $\Upsi^a$ say,
consitituting a reduced field space with the non trivial geometry 
characterised by the intrinsic metric $\GG_{ij} . $ This means that to 
linear order in the deviations $\Upsi^a$  -- provided they  are chosen 
locally as extrinsic coordinates of geodesic normal type -- the total
kinetic metric of the extended system will be expressible in the form
{\be {\rm d}\llam_{\rm tot}^{\,\,2}=\big(\GG_{ij}+\KG_{ij\,a}\Upsi^a\big)
\,{\rm d}\Phhi^i\,{\rm d}\Phhi^j+\Kdel_{ab}\,{\rm d}\Upsi^a
\,{\rm d}\Upsi^b\, ,\label{exmec}\fe}
where $\Kdel_{ab}$ is just a Kronecker unit matrix and the coefficients 
$\GG_{ij}$ and $\KG_{ij\,a}$ are respectively interpretable as 
components of the first and the second fundamental tensor of the 
submanifold. 

The idea is that form of the total potential is such that it
can be decomposed as a sum
{\be \VV_{\!\rm tot}=\VV_{\!\rm cof}+ \VV_{\!\rm red} \label{confy}\fe}
of a relatively slowly varying residual contribution $\VV_{\!\rm red}$ 
and a much more rapidly varying (and usually more highly symmetric) 
confining contribution given to quadratic order in the neigbourhood of 
the submanifold where $\Upsi^a=0 ,$ by an expression of the form
{\be \VV_{\!\rm cof}=\AA^{-1}_{\,ab}\Upsi^a\Upsi^b/8\,\arepsilon\fe}
where the variables $\AA^{-1}_{\,ab}$ are the inverse
components of a moderate valued positive definite matrix $\AA^{ab}$, 
and $\arepsilon$ is a dimensionless constant that has to be taken to 
be very small so as to get the strongly confining limit.

The value $\VV$ of $\VV_{\!\rm red}$ on the submanifold where $\Upsi^a=0 ,$ 
is all that will be needed for determining the effective behaviour of the 
system whenever the available field energy is insufficient to excite 
significant deviations from this submanifold. Thus, in such a low 
energy limit, the system will be effectively describable just in
terms of the reduced model constituted by the reduced field space 
with the metric $\GG_{ij} .$  

It is of interest to consider the effect on the energy measure 
(\ref{enmeas} of the deviations from the reduced model that will 
occur when the confining parameter $\arepsilon$ is not exactly zero,
in which case, to linear order, the residual potential will be given 
in terms of a set of coefficients $\VV_{\!a}$ by an expression
of the form
{\be \VV_{\!\rm red}=\VV+\VV_{\!a}\Upsi^a\, .\fe}
To linear order in $\arepsilon$ and in the deviations $\Upsi^a$
it can be seen from (\ref{exmec}) that in the extended  model
the total energy measure associated with infinitesimal field 
variations ${\rm d}\Phhi^i ,$ ${\rm d}\Upsi^a ,$ across a wall 
will be given by
{\be {\rm d}\UU_{\!\rm tot}^{\,2}=2 \VV\big(\GG_{ij}
+\KG_{ij\,a}\Upsi^a\! +\GG_{ij}\VV_{\!a}\Upsi^a/\VV\!+
\GG_{ij}\AA^{-1}_{\,ab}\Upsi^a\Upsi^b/8\,\arepsilon\VV\big)
\,{\rm d}\Phhi^i\,{\rm d}\Phhi^j+2 \VV\Kdel_{ab}\,{\rm d}
\Upsi^a\,{\rm d}\Upsi^b\, .\label{firsten}\fe}
It can be seen that the first term will be minimised by taking
the deviation components to be given by the formula
{\be \Upsi^a=-4\arepsilon\VV\AA^{ab}\QQ_b\, ,\hskip 1 cm
\QQ_b=(\KG_{ij\,b}+\GG_{ij}\VV_{\!b}/\VV)
\Phhi^{i\pprime}\Phhi^{j\pprime}\, ,\label{cue}\fe}
in which, as before, the prime is used to denote differentiation with 
respect to the kinetic distance measure ${\rm d}\llam$ in the 
reduced space. For such an energy minimising (and thus equilibrium) 
trajectory, it follows that -- to first order in $\arepsilon$ -- the 
total rate of variation of the energy will be given in terms of its 
limit value $\UU^{\pprime}$ in the reduced model by
{\be \UU_{\!\rm tot}^{\,\pprime}=\UU^{\pprime}
\big(1-\arepsilon\VV\AA^{ab}\QQ_a \QQ_b\big)\, ,  \hskip 1 cm
\UU^{\pprime}=\sqrt{2\VV}\, . \fe}
The corresponding integrated total energy over a path through the wall
will therefore be given to linear order by an expression of the form
{\be \UU_{\!\rm tot}= \UU-\arepsilon \WW\, , \fe}
with
{\be \UU=\int\! \sqrt{2\VV}\,{\rm d}\llam
\, ,\hskip 1 cm \WW=\int\! \AA^{ab}\QQ_a \QQ_b\,\VV
\sqrt{2\VV}\,{\rm d}\llam \, .\label{nonc}\fe}

\bigskip\noindent
{\bf Appendix B: Examples of extended field models.}
\medskip

The pentavac doublet model considered in Section 5 was originally
obtained \cite{C04} in just this way by taking the limit $\arepsilon
\rightarrow 0$ in a broken U(1)$\times$ U(1) model involving a pair 
of complex fields with a with a flat kinetic metric of the standard 
O(4) invariant form
{\be {\rm d}\llam_{\rm tot}^{\,\, 2}=5\eeta^2
({\rm d}\Phhi\,{\rm d}\overline\Phhi+{\rm d}\Pssi\,{\rm d}\overline\Pssi)
\, ,\hskip 1cm \Phhi=|\Phhi|{\rm e}^{i\phhi}\, ,\hskip 1 cm
\Pssi=|\Pssi|{\rm e}^{i\pssi}\, ,\fe}
and with a potential given as a sum of the form (\ref{confy})
with a U(1)$\times$U(I) invariant confining part
{\be \VV_{\!\rm cof}=\frac{\VV_{\sharp}}{4}
\left({ (|\Phhi|^2\!-\!1)^2+(|\Pssi|^2\!-\!1)^2}
\right)\, ,\hskip 1 cm \VV_\sharp=\mm_\sharp^{\,4}\, ,\label{cof}\fe}
together with  a weakly variable residual part given in terms of 
the combinations (\ref{theph}) by
{\be \VV_{\!\rm red}= \frac{\VV_{\!\star}}{4}\left(
|\Phhi|^2|\Pssi|^2({\rm cos}\,\thheta+{\rm cos}\,\chhi)
+2/(1-\varepsilon)\right)\, ,\hskip 1 cm \VV_{\!\star}=\mm_\star^4
=\arepsilon \VV_\sharp\, ,
\label{Resi}\fe}
in which the final constant term has been included to adjust the 
minimum of the total potential to the standard value 
$\VV_{\!\rm tot}=0 . $ The ratio
{\be \arepsilon=\frac{\VV_{\!\star}}{\VV_{\sharp}}=\left(\frac
{\mm_\ast}{\mm_\sharp}\right)^4\fe}
will act as a small symmetry breaking parameter.
In this case the minimum of the confining potential $\VV_{\!\rm cof}$
that will be used as the reduced field potential consists of the toroidal 
locus where $|\Phhi|^2=\Pssi|^2=1$, to whose neighbourhood the field will 
be closely confined when $\mm_\sharp\gg \mm_\star$ unless the available 
field energy density is relatively large compared with $\VV_{\!\star}$. 
This neighbourhood will include the five vacuum positions where 
$\VV_{\!\rm tot}$ vanishes, which happens wherever $ {\rm cos}\,\thheta
={\rm cos}\,\chhi=-1$ on the nearby toroidal locus where 
$|\Phhi|^2=|\Pssi|^2=1/(1-\arepsilon)\, .$  

The same reduced field model and linearised extension
can be more elegantly obtained from a variant 
with a confining potential of the same form (\ref{cof}), namely
{\be \VV_{\!\rm cof}=\frac{\mm_\sharp^{\,4}}{4}
\left({ (|\Phhi|^2\!-\!1)^2+(|\Pssi|^2\!-\!1)^2}\right)\, ,\fe}
but with an adjustment of final term in (\ref{Resi}) so as
to get a residual contribution of the more convenient form
{\be \VV_{\!\rm red}= \frac{\mm_\star^{\,4}}{4}
|\Phhi|^2|\Pssi|^2({\rm cos}\,\thheta\!+\!{\rm cos}\,\chhi\!+\!2)=
\frac{\mm_\star^{\,4}}{2}|\Phhi|^2|\Pssi|^2\left({\rm cos}^2\{\thheta/2\}
+{\rm cos}^2\{\chhi/2\}\right)\, ,\fe}
which is such as to ensure that the vacua occur exactly on the 
submanifold of the reduced manifold -- namely the
locus where $|\Phhi|^2=|\Pssi|^2=1$ -- even for finite values of
the dimensionless ratio $(\mm_\star/\mm_\sharp)^4$.
 
To linear order in $\arepsilon$ and $\Upsi$ these variants are 
effectively equivalent, as they are both expressible in the same 
standard form of the kind used in (\ref{exmec}) and (\ref{firsten}) 
by taking the coordinates of the reduced system to be
{\be \Phhi^{_1}=\sqrt 5\,\eeta\,\phhi\, ,\hskip 1 cm \Phhi^{_2}=
\sqrt 5\,\eeta\,\pssi\fe}
and by taking the deviation coordinates to be given by 
{\be \Upsi^{_1}=\sqrt 5\,\eeta\,(|\Phhi|-1)\, ,\hskip 1 cm 
\Upsi^{_2}=\sqrt 5\,\eeta\, (|\Pssi|-1)\, ,\fe}
which gives $\VV_{_1}=\VV_{_2}=2\VV/\sqrt 5\,\eeta\, .$
This means that the nonvanishing components of the first fundamental
tensor on the reduced submanifold (where $\Upsi^{_1}=\Upsi^{_2}=0$) will 
be given by $\GG_{_{11}}=\GG_{_{22}}=1 ,$ while the 
nonvanishing components of the corresponding  second fundamental tensor 
will be given by $\KK_{_{11\,1}}=\KK_{_{22\,2}}=2/\sqrt 5\,\eeta .$
According to (\ref{cue}) this will give
$\QQ_{_1}=2 \sqrt 5\,\eeta\,(2\phhi^{\pprime\,2}+\pssi^{\pprime\,2}) , $ 
$\QQ_{_2}=2 \sqrt 5\,\eeta\,(\phhi^{\pprime\,2}+2\pssi^{\pprime\,2})\, .$
Since the only nonvanishing components of the matrix characterising the 
confining potential will be given by $\AA^{_{11}}=\AA^{_{22}}=
5\eeta^2/8\VV_{\!\star}\, ,\, $ one finally obtains the formula
{\be \AA^{ab}\QQ_a \QQ_b=(\Phhi^{_1\pprime\,4}+\Phhi^{_2\pprime\,4}+4)
/2\VV_{\!\star}\, , \, \label{AQQ}\fe} in which it is to be remarked that 
the variation rates are subject to the normalisation condition
$\Phhi^{_1\pprime\,2}+\Phhi^{_2\pprime\,2}=1\, .$

For a wall of the composite kind obtained by taking  $\chhi=\thheta$, 
one obtains $\VV=\VV_{\!\star}\,{\rm cos}^2\{\thheta/2\}$ so it can be 
seen from (\ref{nonc}) that the ratio of the corresponding values 
$\WW_{\rm I\!I}$ and $\UU_{\rm I\!I}$ will be given by
{\be \WW_{\rm I\!I}/\UU_{\rm I\!I}=2\VV_{\!\star}
\AA^{ab}\QQ_a \QQ_b/3\, ,\fe} whereas for a simple wall the average value 
of $\VV$ will only be half as much so one obtains
{\be \WW_{\rm I}/\UU_{\rm I}=\VV_{\!\star}\AA^{ab}\QQ_a \QQ_b/3\, .\fe}
In the latter case as exemplified by the trajectory $\phhi=2\pssi$
obtained by setting $\chhi=0$ (AB in Figure 2) the variation rates of 
the independent angles $\phhi$ and $\pssi$ will have a two to one ratio, 
so (\ref{AQQ}) will give the value $ \AA^{ab}\QQ_a \QQ_b=117/50 
\VV_{\!\ast}\, ,\,$ from which one finally obtains
{\be \WW_{\rm I}/\UU_{\rm I}=117/150\, .\fe}
In the case of a composite wall as exemplified by the trajectory 
$\phhi=\pssi/3 $ obtained by setting $\chhi=\thheta $ (AC in Figure 2) 
the corresponding ratio will be one to three, which leads
to the value  $ \AA^{ab}\QQ_a \QQ_b=241/100 \VV_{\!\ast}\, ,\,$ from
which one finally obtains 
{\be \WW_{\rm I\!I}/\UU_{\rm I\!I} =241/150\, .\fe}
The ratio of the corresponding net wall tensions,
$\TT_{\rm I}=\UU_{\rm I}-\arepsilon \WW_{\rm I}$ and 
$\TT_{\rm I\!I}=\UU_{\rm I\!I}-\arepsilon \WW_{\rm I\!I} , $ can 
thereby be seen to be given to first order by
{\be \frac{\TT_{\rm I\!I}}{2 \TT_{\rm I}}=1-\frac{62\,\arepsilon}{75} 
\, .\fe}
It can be seen from this by  (\ref{ancrit})  that in such a pentavac
model the simple X junctions will be stable so long as the intersection 
occurs with an acute angle $\aalpha$ that is not too small to satisfy 
the condition that is given in the small $\arepsilon$ limit by
{\be\aalpha>\aalpha_{\rm c}\, ,\hskip 1 cm  \aalpha_{\rm c}
=4\sqrt{\frac {31\,\arepsilon}{75}}\approx 2.57\sqrt\arepsilon
\, ,\label{pental}\fe}
but destablisation can be expected to occur for moderately large
values of $\aalpha$ in cases of the  weakly confined (and more 
numerically tractable)  kind  -- to which the work of Avelino et al
\cite{AMMM06} was restricted -- for which $\arepsilon$ is comparable 
with unity

Such destabilisation by deconfinement does not occur for the
analogous generalisation of category of monovac triplet models of 
the kind considered in Section 6, which can be similarly obtained as a 
low energy limit, for 
$\mm_\ast\ll \mm_\sharp$, from broken U(1)$\times$U(1)$\times$U(1) 
models with a flat kinetic metric of the standard 0(6) invariant form
{\be {\rm d}\llam^2=\eeta^2 ({\rm d}{\Ttheta}\,{\rm d}
\overline{\Ttheta}+{\rm d}{\XX}\,{\rm d}\overline{\XX}
+{\rm d}{\Ssigma}\,{\rm d}\overline{\Ssigma})
\, ,\label{kineb}\fe}
for
{\be  {\Ttheta}=|{\Ttheta}|\,{\rm e}^{i\thheta}
\, ,\hskip 1cm {\XX}=|{\XX}|\,{\rm e}^{i\chhi}\, ,\hskip 1 cm
{\Ssigma}=|{\Ssigma}|\,{\rm e}^{i\ssigma}
\, ,\fe}
with a confining term of the form
{\be \VV_{\rm cof}=\frac{\mm_\sharp^{\,4}}{4}
\left( (|\Ttheta|^2\!-\!1)^2+(|\XX|^2\!-\!1)^2+
(|{\Ssigma}|^2\!-\!1)^2\right)\, ,\label{poteb}\fe}
The required form (\ref{betam}) of the potential $V$ for the ensuing 
reduced model is then obtained by imposing the unimodular retriction
on a weakly variable residual part that is taken to be given in 
terms of a fixed index $\beta$ by
{\be \VV_{\rm red}= 
\frac{\mm_\star^{\,4}}{2}\left(|{\Ttheta}|^2{\rm cos}^2\{\thheta/2\}
+|\XX|^2{\rm cos}^2\{\chhi/2\}+|{\Ssigma}|^2{\rm cos}^2\{\ssigma/2\}
\right)^\beta\, ,\label{betaW}\fe}
so that its dependence on the field moduli will be quartic for 
$\beta=2 ,$ and quadratic in the case $\beta=1$.
It is to be observed that in the special case $\beta=1$
this model will be separable, not just in the reduced limit but also 
(unlike the pentavac case) for finite values of the confinement 
parameter $\arepsilon$, so that it is obvious that its X junctions
will remain stable even for infinitesimally small values of their
acute intersection angle $\aalpha$, and it can easily
be checked directly that there will be a cancellation whereby
the relevant analogue of the pentavac limit formula (\ref{pental})  
works out in this case this case simply to be $\, \aalpha_{\rm c}=0 \, .$

An intrinsically equivalent reduced model is obtainable from an 
algebraically simpler alternative, albeit with not just one but 
8 distinct vacuua, that is specifiable in terms of 6 real variables, 
${\calX}_{\rm i} ,$ ${\calY}_{\rm i} ,$ for i=1, 2, 3, which  
combine as a set of 3 complex variables, whose phases determine
the corresponding angles $\thheta=2\zzeta_1 ,$ $\chhi=2\zzeta_2 ,$ 
$\ssigma=2\zzeta_3 ,$ in the form
{\be{\calX}_{\rm i}+i{\calY}_{\rm i}= {\calZ}_{\rm i}=
|{\calZ}_{\rm i}|\,{\rm e}^{i\,\zzeta_{\rm i}}\, ,\fe}
with the potential contributions in  (\ref{confy}) given 
(in a form that is purely quartic for $\beta=2$) by
{\be \VV_{\rm cof}=\frac{\mm_\sharp^{\,4}}{4}\sum_{\rm i} ( 
{\calX}_{\rm i}^{\, 2}+{\calY}_{\rm i}^{\,2}-1)^2 \ ,\hskip 1 cm
 \VV_{\rm red}= \frac{\mm_\star^{\,4}}{2}\Big(\sum_{\rm i}
{\calX}_{\rm i}^{\, 2}\Big)^\beta \, ,\fe}
so that unimodularity in the low energy limit is obtained in the same
way as before on the assumption that $\mm_\sharp\gg \mm_\star .$
Provided the flat kinetic metric is taken to have the form
{\be {\rm d}\llam^2=4\eeta^2\sum_{\rm i}({\rm d}
{\calX}_{\rm i}^{\, 2}+{\rm d}{\calY}_{\rm i}^{\,2}) \, ,\fe}
this will give a reduced model that will be physically equivalent
to the monovac model described above -- as can be seen by
making the identifications
${\calX}_{_1}={\rm cos}\{\thheta/2\} ,$ 
${\calX}_{_2}={\rm cos}\{\chhi/2\} ,$
${\calX}_{_3}={\rm cos}\{\ssigma/3\} ,$ in the unimodular
limit $|{\calZ}_{_1}|=|{\calZ}_{_2}|=|{\calZ}_{_3}|=1 ,$
which leads to expressions of exactly the same form (\ref{kinet}),
(\ref{betam}) as before. By setting 
$ \phhi=\zzeta_{_1}\, ,\pssi= \zzeta_{_2}\, ,\xsi= \zzeta_{_3}\, ,$ 
it can be seen that the reduced model obtained in this way is 
mathematically equivalent to the octovac model characterised by 
(\ref{doub}). The octovac triplet model differs in principle from the 
corresponding monovac triplet model in that two kinks not just one 
are needed to obtain a complete winding. However this difference would 
not be perceptible in practice within the restricted framework of the 
reduced model, as there would be no way of  telling whether the 
periodicity of the phase variables should be $2\pi$ or $4\pi ,$ and 
therefore no way of knowing whether the relevant vacuua were really 
physically distinct or not.

\end{document}